\documentclass{aa}
\usepackage{graphicx}
\newcommand{\lsim}{\raisebox{-5pt}{$\;\stackrel{\textstyle <}{\sim}\;$}}
\newcommand{\gsim}{\raisebox{-5pt}{$\;\stackrel{\textstyle >}{\sim}\;$}}

\begin{document}
   \title{ The early build-up of dust in galaxies: \\
   A study of Damped Ly $\alpha$ Systems
 }
 
 \titlerunning{The early build-up of dust in galaxies}

   \author{G. Vladilo }

   \offprints{G. Vladilo}

   \institute{
              Osservatorio Astronomico di Trieste - Istituto Nazionale di Astrofisica, 
              Via G.B. Tiepolo 11, 34131 Trieste, Italy
              \email{vladilo@ts.astro.it}
                              }

   \date{Received ...; accepted ...}

 \abstract{
We present a study of the early build-up of dust in high redshift galaxies.
The study is based on the analysis of  38  Damped  Ly$\alpha$ systems (DLAs)
for which we derive  
the fraction of iron atoms in dust form, $f_\mathrm{Fe}$.
The sample is representative of metal-poor
galaxies in the redshift range $ 0.6 \leq z \leq 3.4$
selected on the basis of their absorption \ion{H}{i} column density
($N(\ion{H}{i})  \geq  2 \times 10^{20}$ atoms cm$^{-2}$).  
We find that the dust fraction  
increases with metallicity,   from  $f_\mathrm{Fe} \simeq 0$ 
at [Fe/H]$\sim -2$ dex, up to $f_\mathrm{Fe} \simeq 0.9$ 
at solar metallicity;  
the increase is fast below [Fe/H] $\simeq -1$ dex and mild at higher metallicities. 
We also find some evidence for an increase of $f_\mathrm{Fe}$ with
cosmic time; 
a large fraction of the systems  younger than $\approx 3$ Gyr
has $f_\mathrm{Fe} \la 0.5$.   
These results indicate the dust-to-metal ratio increases
in the course of chemical evolution, at variance with 
the hypothesis of an approximately constant dust-to-metal ratio,
commonly adopted in models of galactic evolution. 
This hypothesis is consistent with local and high-redshift data  
only  when the   metallicity is relatively high  ([Fe/H]$\ga -1$ dex). 
The  results of this work suggest that the main mechanisms of dust
formation  may be rather sensitive to the level of 
metallicity attained by a galaxy in the course of its chemical evolution. 
A metallicity-dependent dust production by SNe II seems to be
the most promising mechanism for explaining the rise of $f_\mathrm{Fe}$
at [Fe/H] $\la -1$ dex. 
}

   \maketitle


\section{Introduction}

Interstellar dust plays a key role in a variety of astrophysical processes
relevant to galactic evolution and affects, with its presence, many measurements of basic physical properties
of stars and galaxies. %
 For these reasons, 
 understanding the process of dust formation and evolution on a galactic scale 
 is of great interest in several areas of
astrophysical research.  
In the course of galactic evolution 
the abundance of
dust is expected to track that of the metals synthesized and ejected  from  the stars. 
Therefore, the dust-to-gas ratio,
$\cal D$,  is expected to correlate   with the metallicity, $\cal Z$
or, in other words,
the fraction of heavy elements in dust form, $f_\mathrm{d} \approx {\cal D/Z}$, 
is expected to stay 
approximately constant  (e.g. Franco \& Cox 1986).
In the local universe the existence of a   $\cal{Z - D }$ correlation  
is supported by empirical evidence.  
For instance, Issa et al. (1990) found a  good 
correlation 
for  6 spirals of the Local Group and the Magellanic Clouds,
estimating $\cal D$  
from the visual extinction per hydrogen atom. 
Schmidt \& Boller (1993) found 
a similar correlation  
for 23 local dwarf galaxies estimating $\cal D$  
 from  IRAS dust mass and \ion{H}{i} integrated-emission mass.
From a similar analysis
Lisenfeld \& Ferrara (1998)   found a good 
correlation for 28 dwarf irregular galaxies,
but not for 16 blue compact dwarfs; they attribute these differences
to variations of the star-formation   and mass-outflow rates
between these different types of galaxies.  

The empirical $\cal{Z - D }$ correlation 
found in the local universe is, to date, the main
observational input to models of galactic evolution which incorporate
the dust component. 
A self-consistent treatment of the metal and dust content
in the models 
requires taking into account
the processes of grain formation, accretion and destruction
(e.g. Dwek \& Scalo 1980), increasing
in such a way the number of the 
parameters with few, if any,
observational constraints. 
 The detailed treatment of all these processes can be 
performed using numerical models.
Following this approach, and requiring that accretion and destruction
time-scales evolve in a similar fashion, 
Dwek (1998)  found that the interstellar ${\cal D/Z}$ mass ratio  
is  approximately constant in the Milky Way.
 %
More elementary, analytical models  
have been recently proposed  to reproduce the dust cycle in galaxies
(e.g. Hirashita 1999, Edmunds 2001, Inoue 2003). 
An approximately constant ${\cal D/Z}$ mass ratio is generally predicted
by these  studies, with the exception of Inoue (2003),
who proposes models with increasing  ${\cal D/Z}$    in the course of evolution. 
The observed linear  $\cal{Z - D }$ relation in the local universe
 is interpreted by Inoue as a sequence
of constant galactic age.
Measuring the dust-to-metal ratios in galaxies
caught in their early stages of evolution is fundamental
to probe the validity of these models.
This can be done by
studying the dust content of high-redshift galaxies, a field of research that 
is receiving a special attention given its potential impact on our understanding of the
early universe 
(e.g. Silva et al. 1998, Hirashita \& Ferrara 2002, Morgan \& Edmunds 2003).   
 
Quasar absorption-line systems 
can be used to investigate the relation between dust   and metallicity
at high redshift on observational grounds.  This is true for
the Damped Ly $\alpha$ (DLA) systems,
the class of absorbers with the highest \ion{H}{I} column density
($N(\ion{H}{I}) > 10^{20.3}$ atoms cm$^{-2}$),
which are believed to originate in galaxies  located along the quasar line of sight
(Wolfe et al. 1986). 
Measuring the metallicity and the dust abundance
of DLA systems makes it possible  to probe the  ${\cal D - Z}$ relation
in primeval galaxies. 
The metallicity  is determined from  column-density measurements,
which can be rather accurate  (e.g. Lu et al. 1996, Prochaska \& Wolfe 1999,
Molaro et al. 2000, Ledoux et al. 2003).
Element abundances do not suffer from severe ionization corrections in DLA systems
(e.g. Vladilo et al. 2001),
and the metallicity level can be estimated pretty well, at least for elements
not affected by dust depletion (Pettini et al. 1999).  
Measuring the  dust abundance, however, is less straightforward. 

In principle, the amount of dust   in DLA systems  can be estimated from 
the reddening  of
the  background quasar. 
This technique, however, cannot be applied to individual
 systems given the
uncertain, variable continuum of the quasar. 
As a consequence, quasar extinction
has only been estimated  statistically, by  comparing samples with and
 without foreground DLAs (Pei et al. 1991). 
The dust-to-gas ratio in DLA systems estimated in this way
is between 5\% and 20\% of the Galactic value, in broad agreement
with the low level of metallicity of DLA systems.
However, this type of analysis does not 
give information on the evolution of the dust abundance with
the metallicity. 
 Also the dust emission properties
 cannot be employed to derive the dust abundance in DLA systems.
In fact, attempts to identify the emission of
intervening galaxies in the quasar field have been successful only for a dozen
of absorbers at $z \la 1$ (Le Brun et al. 1997, Turnshek et al. 2001; see also
Kanekar \& Chengalur 2003), but not in 
the redshift interval where DLAs are most commonly identified 
and investigated ($z \ga 2$). 

So far, the only method to estimate the dust abundance of individual DLAs relies  
on the comparison of the elemental abundances of refractory and non-refractory
elements. Since the pioneer investigation on Cr and Zn by Pettini et al. (1994), 
the evidence for differential depletion of refractory elements
relative to the volatile element Zn has accumulated over the years and
is now well established  (Hou et al. 2001, Prochaska \& Wolfe 2002).
An additional argument supporting the connection between depletion and dust
is the correlation between depletion and fraction of molecular hydrogen  
(Levshakov et al. 2000, Ledoux et al. 2003). 
 
The existence of a correlation between differential depletion 
and metallicity, reported by Ledoux et al. (2003),
suggests that the dust content of DLA systems   evolves in the course
of chemical evolution. However,
in order to study the evolutionary properties of the dust we need
to convert the differential depletions into dust-to-gas or dust-to-metal ratios. 
%
In a previous work we presented a method for estimating 
these ratios in DLA systems,
taking also into account the depletion of Zn self-consistently
(Vladilo 1998; Paper I). 
In that work we derived
a $\cal{Z - D }$ correlation  between the
metallicity corrected for dust effects and the dust-to-gas ratio.
These two quantities, however, are not independent when the
observed metallicity is very low. 
A much safer approach consists in studying the evolution
of the {\em dust-to-metal ratio}, rather than that of the {\em dust-to-gas ratio}.
The former  is derived from the column densities of two
metals and
is   less affected by error propagation than 
the latter, which requires, in addition,
the use of the \ion{H}{i} column density.  

A systematic study of the evolution 
of the dust-to-metal ratio in DLA systems is still missing. 
This type of study requires converting the differential depletions
measured at face value into the   
{\em fraction of metals in dust form}, $f_d$, taking
into account (i) variations of the dust composition that may occur in different types
of galactic environments (e.g. galaxies with different metallicities)
and (ii) nucleosynthetic effects that may affect the 
volatile/refratory abundance ratio (e.g. evolution of
the Zn/Fe ratio). 
 In a recent work we presented a  
methodology aimed at tackling these problems 
(Vladilo 2002a; Paper II).
This methodology was successfully applied to recover 
(i) the
(stellar) abundance pattern of the Small Magellanic Cloud
starting from interstellar measurements affected by dust depletion;
(ii) the $\alpha$/Fe ratio in DLA systems corrected for dust effects
(Vladilo 2002b; Paper III).  

In the present paper we   apply this  new method to derive
the fraction of iron in dust form, $f_\mathrm{Fe}$,
for   the   sample of DLA systems
with measurements of the Zn/Fe ratio.
This sample now includes 38 absorbers, 
more than twice the sample of Paper I. 
We use the results of the most recent studies of Zn abundance
in metal-poor stars 
to derive an educated guess for 
the intrinsic Zn/Fe ratio in DLA systems which, at variance with Paper I,
 is now a free parameter of the method. 
Finally, for the first time in this type of investigation,  the robustness of the results 
is tested against possible bias   introduced by the observations
or by the method itself. 
The paper is organized as follows. 
In the next Section we present the method of analysis and the sample. 
 In Section 3 we present the results and test
 their robustness. In Section 4 we discuss the implications of the
 results in terms of formation and evolution of dust in primeval
 galaxies. The conclusions are summarized in Section 5.

\section{Analysis of the observational data}

\subsection{ The method }

In order to    study the evolutionary properties of  
dust  in DLA systems we   measure the fraction of metals in dust form
in each system. 
We call {\em dust fraction}  of an element X the number ratio 
\begin{equation}
\label{Fraction}
 f_\mathrm{X} 
= N_\mathrm{dust}(\mathrm{X})/N_\mathrm{tot}(\mathrm{X}) ~,
\end{equation} 
where  $N_\mathrm{dust}(\mathrm{X})$  and $N_\mathrm{tot}(\mathrm{X})$
are the column densities of X (atoms cm$^{-2}$) 
in dust form and in the medium (gas plus dust), respectively.
{\em The dust fraction of a metal} indicates how efficient the ISM of a galaxy is
to incorporate  part of its heavy elements into dust grains and 
{\em represents, in practice, a dust-to-metal ratio by number}.
Here we adopt the dust fraction of iron, $f_\mathrm{Fe}$, as an estimator of 
  the dust-to-metals ratio  in DLA systems.
 Iron is a natural choice since it
  is easy to measure
in DLAs and is a reference element used in different types of abundance studies.
Most important, iron is a refractory element which traces  the dust  
even in the harshest interstellar conditions, where dust grains tend
to be destroyed. This is demonstrated by the fact that  in the warm halo
gas of the Milky Way, where the level of depletions is known to be generally low,
the dust fraction   of iron is still quite high,
$f_{\mathrm{Fe},wh} \simeq 0.77$  (Savage \& Sembach 1996). 

%

\subsubsection{ Derivation of  the dust fraction of iron }

The dust fraction   of iron 
 can be derived by comparing the observed
 column density of iron with that of a volatile element X 
if  we can make an educated guess of 
the intrinsic abundance ratio X/Fe in the   DLA system. 
The method works best when the differential
dust depletion between Fe and X is high. The most natural choice
for X is Zn, a volatile element that tracks Fe well 
 in the course of nucleosynthetic evolution  (see Section 2.2.3).
If we call [Zn/Fe]$_g$ the educated guess of the intrinsic ratio in DLAs,
we can use the equation 
\begin{equation}
\label{Rho}
\varrho
- { f_{\mathrm{Zn},i} 10^{[ \mathrm{Zn \over Fe} ]_\mathrm{g}   \varepsilon_\mathrm{Zn} }
\over 
f_{\mathrm{Fe},i} 10^{[ \mathrm{Zn \over Fe} ]}
}
\varrho^{(1+\eta_\mathrm{Zn})} 
+ { 10^{[ \mathrm{Zn \over Fe} ]_\mathrm{g}  } - 10^{[ \mathrm{Zn \over Fe} ]}
\over
f_{\mathrm{Fe},i} 10^{[ \mathrm{Zn \over Fe} ]}
 } 
=0 ~~,
\end{equation} 
derived in Paper III
to obtain the  normalized dust-to-metal ratio 
$\varrho = f_{\mathrm{Fe}}/f_{\mathrm{Fe},i}$
from the ratio [Zn/Fe]  measured in the DLA.\footnote
{
[Zn/Fe] $\equiv \log \{ N(\mathrm{Zn})/N(\mathrm{Fe}) \} - \log (\mathrm{Zn/Fe})_{\sun}$,
where $ N(\mathrm{Zn})$ and $N(\mathrm{Fe})$ are the   column densities
derived from spectroscopic measurements. The reference
solar abundance by number, $\log (\mathrm{Zn/Fe})_{\sun} = -2.86$,  is taken
from   meteoritic measurements  (Anders \& Grevesse 1989).
} 
In the above equation
the fractions in dust $f_{\mathrm{Fe},i}$ and $f_{\mathrm{Zn},i}$
are measured in  one particular phase $i$  
of the Galactic ISM (e.g. cold gas, or warm gas in the disk or in the halo).
{\em Without loss of generality}, we adopt here the warm gas in the disk as a reference
value: we stress that, at variance with Paper I, the present method
allows for variations of the dust composition
among different interstellar phases (see Paper II). 
These variations   are taken into account
by the parameter $\eta_\mathrm{Zn}$, which
describes how the 
  Zn/Fe ratio in the dust varies as a function of $\varrho$.
The parameter $\eta_\mathrm{Zn}$ is determined
 from studies of the Galactic depletion patterns (Paper II).
 The method also allows the dust composition to vary as a function of the
 composition of the medium. This dependence is accounted for by 
 the parameter $\varepsilon_\mathrm{Zn}$,     discussed below.
Finally, we stress that Eq. \ref{Rho}  takes into
account in a self-consistent way
the fact that also Zn is partly depleted, while dust depletion studies
in literature ignore this effect. 
  
Eq. \ref{Rho}   is a transcendent equation in the unknown $\varrho$
that we solve by iterations. The equation
  has  a unique solution inside     
the   interval   $0 < \varrho \leq \varrho_\mathrm{lim}$,
where $\varrho_\mathrm{lim}$ is the maximum possible value of $\varrho$,
corresponding to a fraction in dust of iron of 100\%. Solutions outside
this interval are not physical.

\subsubsection{ Derivation of the metallicity }

Once the fraction in dust of iron is known, it is straightforward to derive
the total column density of iron (gas plus dust) 
and from this
the iron abundance corrected for dust effects, 
[Fe/H]$_\mathrm{c}$, used here as the indicator of metallicity.
In most cases 
the dust-corrected metallicity [Fe/H]$_\mathrm{c}$  is very similar
to the zinc abundance [Zn/H] taken at face value, which is commonly
adopted in literature as the indicator of metallicity in DLA systems, with
a typical difference [Fe/H]$_\mathrm{c} -$ [Zn/H]$\lsim  +0.1$ dex.

 \subsection{ The parameters }
 
\subsubsection{ The dust fraction  of zinc in the Galactic ISM }

Local interstellar studies indicate that a fraction of zinc is 
incorporated in dust form even in lines of sight with moderate level of depletion.
In Paper II we derived
the fraction in dust of zinc  in the Galactic warm disk gas, 
$f_{\mathrm{Zn},{wd}}=0.59$, 
using the same lines of sight  with accurate {\em HST} measurements
adopted by Savage \& Sembach (1996) in their study of depletion patterns.
Unfortunately, that sample of lines of sight included only a few zinc
determinations. For this reason, we now derive 
$f_{\mathrm{Zn},{wd}}$ also using the largest compilation of  
{\em HST} measurements of zinc and chromium
(Roth \& Blades 1995). As a result,
we obtain $f_{\mathrm{Zn},{wd}}=0.30$
using the same scaling law of interstellar depletions adopted in Paper II
and imposing that the dust-corrected abundance of Cr in the local ISM
is solar.  
We consider that the range   $0.30 \leq f_{\mathrm{Zn},{wd}} \leq 0.59$
is   fairly representative of the uncertainty of this parameter.

\subsubsection{ The parameter $\varepsilon_\mathrm{Zn}$ }

The parameter    $\varepsilon_\mathrm{Zn}$ describes how  
the Zn/Fe ratio in the dust  composition varies  as a function of the same ratio
in the composition of the medium.
By definition,  $\varepsilon_\mathrm{Zn} = 1$ if a percent
variation of the Zn/Fe ratio in the medium yields
the same percent variation in the dust (Paper II). 
If, on the other hand, the Zn/Fe ratio in the dust is totally
independent of variations of the ratio in the medium, then
$\varepsilon_\mathrm{Zn} = 0$. 
In principle,   $\varepsilon_\mathrm{Zn}$
can be derived studying depletion patterns in galaxies
with a known chemical composition. 
A preliminary study of SMC data is consistent with
$\varepsilon_\mathrm{Zn} \simeq 1$.
Here we consider both cases $\varepsilon_\mathrm{Zn} = 1$
and $\varepsilon_\mathrm{Zn} = 0$ as extreme possibilities
to solve Eq. \ref{Rho} starting from reference Galactic ISM values. 
It is important to note that, given the form of Eq. \ref{Rho},
in which $\varepsilon_\mathrm{Zn}$
multiplies [Zn/Fe]$_\mathrm{g}$,
{\em the choice of $\varepsilon_\mathrm{Zn}$  is not critical
if the intrinsic Zn/Fe ratio is approximately solar}. 
Luckily, the Zn/Fe is indeed close to solar and 
this explains why the results presented below are rather
independent of the choice of $\varepsilon_\mathrm{Zn}$. 

\subsubsection{ The Zn/Fe ratio in DLA systems}

In order to make an educated guess of
the Zn/Fe ratio in DLA systems  we must rely on abundance
studies of metal-poor stars. 
The early studies   of  Galactic  stars 
yielded a Zn/Fe ratio close to solar
(Sneden et al. 1991), but some recent work indicate that 
the Zn/Fe ratio can be overabundant  at very low metallicity
 (Primas et al. 2000; Prochaska \& Wolfe, 2002).
 The latter results have been used to support
the notion that  Zn/Fe ratio may decrease in the course of chemical evolution
owing to a  different nucleosynthetic origin of Zn and Fe (see e.g.
Umeda \& Nomoto, 2002). 

A large amount of Zn abundance measurements have been recently
published by three different groups, all indicating
that {\em the Zn/Fe ratio is very close to solar, or slightly enhanced,
in the range of metallicities typical of DLA systems}
(Mishenina et al. 2002, Gratton et al. 2003, Nissen et al. 2003).
The 87 measurements performed by Mishenina et al., derived from
  spectra collected at the 1.93-m telescope
of Haute Provence, 
yield  a mean value
$<$[Zn/Fe]$>$ $ =+0.08 \pm 0.13$ dex (1$\sigma$)
in the interval $-3 \leq \mathrm{[Fe/H]} \leq -0.2$. 
 The 48 measurements obtained from UVES/VLT spectra
 by Gratton et al. 
yield 
$<$[Zn/Fe]$>$ $ =+0.13 \pm 0.13$ dex
in the interval $-2.4 \leq \mathrm{[Fe/H]} \leq -0.3$.
The 29 measurements by Nissen et al., 
also obtained with UVES/VLT data, yield 
$<$[Zn/Fe]$>$ $ =+0.04 \pm 0.08$ dex
in the interval $-2.4 \leq \mathrm{[Fe/H]} \leq -0.7$.
The scales of the zinc abundances in these measurements
might be in error up to 0.1 dex owing to differences between
the photospheric and meteoritic solar abundance. For instance,
Gratton et al.  used 
a reference solar ratio  
$\log (\mathrm{Zn/Fe})_{\sun} = -2.95$
derived from a consistent analysis of the
solar spectrum. Using instead the 
solar meteoritic value $\log (\mathrm{Zn/Fe})_{\sun} = -2.86$
(Anders \& Grevesse 1989), one would derive
$<$[Zn/Fe]$>$ $ =+0.04 \pm 0.13$ dex from Gratton et al.'s data. 
Also a systematic difference in the  scale 
of effective temperature 
could easily account for the    
mean offset relative to  the solar value. 
 
The [Zn/Fe] data 
{\em do not show a trend with metallicity}, 
with the possible exception of a very weak decrease with [Fe/H]
in the Nissen et al.'s sample and in the sub-sample of
halo stars of Mishenina et al., neither
statistically significant.  Therefore
a constant value of [Zn/Fe]$_\mathrm{g}$
is a reasonable choice 
in the metallicity interval typical of DLA systems. 
We considered two possible values, namely  
  [Zn/Fe]$_\mathrm{g}=0$, representative of the solar value,
and  [Zn/Fe]$_\mathrm{g}=+0.1$, representative  
of a modest enhancement.   
The potential effects of an hypothetic decrease of Zn/Fe with Fe/H
are considered in the discussion.

\begin{center} 
\begin{table*} 
\scriptsize{
\caption{  Fraction of iron in dust, $f_\mathrm{Fe}$, in Damped Ly $\alpha$ systems estimated for different sets of input parameters$^a$. }
\begin{tabular}{lcllllllll}
\hline \hline
Identifier &  $z$ & $\log N(\ion{H}{i})^b$ & $\log N(\ion{Fe}{ii})^b$ & $\log N(\ion{Zn}{ii})^b$ 
& References &  $f_\mathrm{Fe}$ (Sa)$^c$ &  $f_\mathrm{Fe}$ (Sb)$^d$ &  $f_\mathrm{Fe}$ (E0)$^e$ &   $f_\mathrm{Fe}$ (E1)$^f$   \\
\hline
0000-263   &   3.3901 & 21.41$\pm$0.08 & 14.77$\pm$0.03 & 12.01$\pm$0.05 & 13    & $0.215^{+0.100}_{-0.113}$ & $0.215^{+0.099}_{-0.113}$& $0.011^{+0.124}_{-0.011}$ & $0.011^{+0.124}_{-0.011}$ \\
0013-004   &   1.9731 & 20.70$\pm$0.05 & 14.37$\pm$0.01 & 12.63$\pm$0.06 & 6,16  & $0.979^{+0.004}_{-0.005}$ & $0.948^{+0.008}_{-0.009}$& $0.953^{+0.008}_{-0.010}$ & $0.971^{+0.006}_{-0.007}$ \\
0058+019   &   0.6125 & 20.08$\pm$0.15 & 15.24$\pm$0.04 & 12.81$\pm$0.13 & 18    & $0.669^{+0.132}_{-0.165}$ & $0.646^{+0.111}_{-0.149}$& $0.547^{+0.147}_{-0.186}$ & $0.552^{+0.156}_{-0.190}$ \\
0058-2914  &   2.6711 & 21.20$\pm$0.15 & 14.96$\pm$0.14 & 13.04$\pm$0.15 & 2     & $0.961^{+0.020}_{-0.054}$ & $0.916^{+0.036}_{-0.066}$& $0.915^{+0.041}_{-0.090}$ & $0.941^{+0.032}_{-0.089}$ \\
0100+130   &   2.3090 & 21.40$\pm$0.05 & 14.99$\pm$0.04 & 12.45$\pm$0.02 & 24    & $0.534^{+0.056}_{-0.060}$ & $0.525^{+0.052}_{-0.056}$& $0.396^{+0.064}_{-0.070}$ & $0.397^{+0.065}_{-0.070}$ \\
0149+33    &   2.1400 & 20.50$\pm$0.10 & 14.23$\pm$0.02 & 11.50$\pm$0.10 & 24    & $0.252^{+0.166}_{-0.204}$ & $0.252^{+0.163}_{-0.204}$& $0.058^{+0.202}_{-0.058}$ & $0.058^{+0.202}_{-0.058}$ \\
0201+365   &   2.4620 & 20.38$\pm$0.04 & 15.01$\pm$0.00 & 12.76$\pm$0.05 & 20,22 & $0.839^{+0.032}_{-0.038}$ & $0.788^{+0.028}_{-0.032}$& $0.738^{+0.039}_{-0.044}$ & $0.756^{+0.044}_{-0.048}$ \\
0302-223   &   1.0094 & 20.36$\pm$0.11 & 14.67$\pm$0.04 & 12.45$\pm$0.04 & 18    & $0.856^{+0.035}_{-0.044}$ & $0.802^{+0.032}_{-0.037}$& $0.758^{+0.045}_{-0.052}$ & $0.778^{+0.050}_{-0.058}$ \\
0347-383   &   3.0250 & 20.56$\pm$0.05 & 14.47$\pm$0.05 & 12.23$\pm$0.12 & 8$^g$ & $0.841^{+0.071}_{-0.111}$ & $0.790^{+0.065}_{-0.092}$& $0.741^{+0.093}_{-0.126}$ & $0.759^{+0.102}_{-0.136}$ \\
0405-443   &   2.5950 & 20.90$\pm$0.10 & 15.09$\pm$0.06 & 12.53$\pm$0.13 & 8$^g$ & $0.517^{+0.167}_{-0.205}$ & $0.509^{+0.150}_{-0.198}$& $0.377^{+0.188}_{-0.245}$ & $0.377^{+0.193}_{-0.245}$ \\
0454+039   &   0.8597 & 20.69$\pm$0.06 & 15.17$\pm$0.04 & 12.33$\pm$0.09 & 18    & $0.045^{+0.192}_{-0.045}$ & $0.045^{+0.191}_{-0.045}$& $0.000^{+0.038}_{-0.000}$ & $0.000^{+0.038}_{-0.000}$ \\
0458-02    &   2.0400 & 21.65$\pm$0.09 & 15.40$\pm$0.05 & 13.13$\pm$0.02 & 24    & $0.821^{+0.037}_{-0.043}$ & $0.773^{+0.031}_{-0.036}$& $0.716^{+0.044}_{-0.049}$ & $0.732^{+0.049}_{-0.054}$ \\
0515-4414  &   1.1510 & 20.45$\pm$0.10 & 14.24$\pm$0.20 & 12.11$\pm$0.04 & 3     & $0.905^{+0.055}_{-0.139}$ & $0.847^{+0.067}_{-0.120}$& $0.822^{+0.091}_{-0.168}$ & $0.849^{+0.091}_{-0.184}$ \\
0528-2505  &   2.8110 & 21.11$\pm$0.04 & 15.47$\pm$0.02 & 13.27$\pm$0.03 & 1     & $0.868^{+0.020}_{-0.023}$ & $0.813^{+0.018}_{-0.020}$& $0.774^{+0.026}_{-0.029}$ & $0.796^{+0.029}_{-0.032}$ \\
0551-3637  &   1.9615 & 20.50$\pm$0.08 & 15.05$\pm$0.05 & 13.02$\pm$0.05 & 7     & $0.939^{+0.015}_{-0.022}$ & $0.885^{+0.021}_{-0.026}$& $0.875^{+0.027}_{-0.036}$ & $0.905^{+0.026}_{-0.037}$ \\
0812+32    &   2.6260 & 21.35$\pm$0.10 & 15.10$\pm$0.03 & 13.04$\pm$0.02 & 21    & $0.930^{+0.009}_{-0.011}$ & $0.874^{+0.011}_{-0.013}$& $0.860^{+0.016}_{-0.017}$ & $0.889^{+0.016}_{-0.018}$ \\
0841+129   &   2.3745 & 21.00$\pm$0.10 & 14.87$\pm$0.04 & 12.20$\pm$0.05 & 1     & $0.357^{+0.094}_{-0.105}$ & $0.356^{+0.091}_{-0.104}$& $0.187^{+0.112}_{-0.129}$ & $0.187^{+0.113}_{-0.129}$ \\
0935+417   &   1.3726 & 20.52$\pm$0.10 & 14.82$\pm$0.10 & 12.25$\pm$0.01 & 12    & $0.500^{+0.124}_{-0.144}$ & $0.493^{+0.114}_{-0.138}$& $0.357^{+0.141}_{-0.170}$ & $0.357^{+0.144}_{-0.171}$ \\
1104-1805  &   1.6616 & 20.85$\pm$0.01 & 14.77$\pm$0.02 & 12.48$\pm$0.01 & 9     & $0.804^{+0.018}_{-0.019}$ & $0.759^{+0.015}_{-0.016}$& $0.697^{+0.020}_{-0.021}$ & $0.711^{+0.022}_{-0.023}$ \\
1117-1329  &   3.3505 & 20.84$\pm$0.08 & 14.82$\pm$0.05 & 12.25$\pm$0.06 & 14    & $0.500^{+0.098}_{-0.110}$ & $0.493^{+0.090}_{-0.105}$& $0.357^{+0.112}_{-0.129}$ & $0.357^{+0.113}_{-0.130}$ \\
1157+0128  &   1.9436 & 21.80$\pm$0.10 & 15.50$\pm$0.07 & 13.09$\pm$0.08 & 18    & $0.690^{+0.102}_{-0.122}$ & $0.664^{+0.085}_{-0.107}$& $0.571^{+0.113}_{-0.136}$ & $0.577^{+0.120}_{-0.140}$ \\
1209+0919  &   2.5835 & 21.40$\pm$0.10 & 15.22$\pm$0.04 & 12.98$\pm$0.05 & 21    & $0.844^{+0.042}_{-0.054}$ & $0.793^{+0.037}_{-0.045}$& $0.744^{+0.053}_{-0.062}$ & $0.763^{+0.059}_{-0.068}$ \\
1210+1731  &   1.8918 & 20.60$\pm$0.10 & 14.97$\pm$0.06 & 12.37$\pm$0.03 & 19    & $0.458^{+0.091}_{-0.101}$ & $0.454^{+0.086}_{-0.098}$& $0.308^{+0.106}_{-0.121}$ & $0.308^{+0.107}_{-0.122}$ \\
1215+33    &   1.9990 & 20.95$\pm$0.07 & 14.75$\pm$0.05 & 12.33$\pm$0.05 & 24    & $0.682^{+0.073}_{-0.081}$ & $0.657^{+0.061}_{-0.071}$& $0.562^{+0.080}_{-0.091}$ & $0.567^{+0.085}_{-0.094}$ \\
1223+178   &   2.4661 & 21.50$\pm$0.10 & 15.16$\pm$0.02 & 12.55$\pm$0.03 & 19,20 & $0.450^{+0.046}_{-0.049}$ & $0.446^{+0.044}_{-0.047}$& $0.298^{+0.054}_{-0.058}$ & $0.298^{+0.055}_{-0.058}$ \\
1253-0228  &   2.7830 & 21.85$\pm$0.20 & 15.36$\pm$0.04 & 12.77$\pm$0.07 & 21    & $0.473^{+0.107}_{-0.121}$ & $0.468^{+0.099}_{-0.117}$& $0.326^{+0.123}_{-0.144}$ & $0.326^{+0.124}_{-0.145}$ \\
1328+307   &   0.6922 & 21.25$\pm$0.02 & 14.95$\pm$0.15 & 12.72$\pm$0.12 & 11    & $0.852^{+0.087}_{-0.172}$ & $0.799^{+0.086}_{-0.144}$& $0.753^{+0.121}_{-0.194}$ & $0.773^{+0.131}_{-0.208}$ \\
1331+170   &   1.7764 & 21.18$\pm$0.04 & 14.62$\pm$0.00 & 12.54$\pm$0.03 & 24    & $0.925^{+0.009}_{-0.010}$ & $0.869^{+0.010}_{-0.011}$& $0.853^{+0.014}_{-0.016}$ & $0.882^{+0.015}_{-0.017}$ \\
1351+318   &   1.1491 & 20.23$\pm$0.10 & 14.74$\pm$0.09 & 12.52$\pm$0.13 & 24    & $0.856^{+0.074}_{-0.132}$ & $0.802^{+0.072}_{-0.110}$& $0.758^{+0.102}_{-0.151}$ & $0.778^{+0.111}_{-0.164}$ \\
1354+258   &   1.4200 & 21.54$\pm$0.06 & 15.03$\pm$0.09 & 12.59$\pm$0.13 & 24    & $0.658^{+0.152}_{-0.196}$ & $0.636^{+0.128}_{-0.179}$& $0.535^{+0.170}_{-0.222}$ & $0.539^{+0.180}_{-0.226}$ \\
1451+1223  &   2.2550 & 20.30$\pm$0.15 & 14.33$\pm$0.07 & 11.85$\pm$0.11 & 4     & $0.612^{+0.139}_{-0.167}$ & $0.596^{+0.119}_{-0.155}$& $0.484^{+0.154}_{-0.192}$ & $0.487^{+0.160}_{-0.194}$ \\
2206-199A  &   1.9200 & 20.65$\pm$0.10 & 15.32$\pm$0.02 & 12.91$\pm$0.01 & 23    & $0.699^{+0.021}_{-0.021}$ & $0.671^{+0.017}_{-0.018}$& $0.580^{+0.023}_{-0.024}$ & $0.586^{+0.024}_{-0.025}$ \\
2230+025   &   1.8642 & 20.85$\pm$0.08 & 15.18$\pm$0.02 & 12.80$\pm$0.03 & 24    & $0.717^{+0.032}_{-0.034}$ & $0.687^{+0.026}_{-0.028}$& $0.601^{+0.035}_{-0.037}$ & $0.608^{+0.037}_{-0.039}$ \\
2231-0015  &   2.0662 & 20.56$\pm$0.10 & 14.66$\pm$0.07 & 12.35$\pm$0.03 & 24    & $0.787^{+0.055}_{-0.065}$ & $0.745^{+0.046}_{-0.054}$& $0.678^{+0.063}_{-0.073}$ & $0.690^{+0.069}_{-0.078}$ \\
2243-6031  &   2.3300 & 20.67$\pm$0.02 & 14.92$\pm$0.03 & 12.22$\pm$0.03 & 10    & $0.310^{+0.067}_{-0.072}$ & $0.309^{+0.066}_{-0.072}$& $0.129^{+0.081}_{-0.089}$ & $0.129^{+0.081}_{-0.089}$ \\
2314-409   &   1.8573 & 20.90$\pm$0.10 & 15.08$\pm$0.10 & 12.52$\pm$0.10 & 5     & $0.513^{+0.168}_{-0.206}$ & $0.506^{+0.151}_{-0.199}$& $0.372^{+0.189}_{-0.246}$ & $0.373^{+0.193}_{-0.247}$ \\
2342+3417  &   2.9084 & 21.10$\pm$0.10 & 14.98$\pm$0.06 & 12.50$\pm$0.11 & 21    & $0.611^{+0.134}_{-0.161}$ & $0.595^{+0.115}_{-0.149}$& $0.483^{+0.149}_{-0.184}$ & $0.486^{+0.155}_{-0.187}$ \\
2359-0216  &   2.0950 & 20.70$\pm$0.10 & 14.54$\pm$0.03 & 12.60$\pm$0.03 & 24    & $0.956^{+0.006}_{-0.007}$ & $0.908^{+0.009}_{-0.011}$& $0.905^{+0.012}_{-0.014}$ & $0.933^{+0.010}_{-0.013}$ \\
\hline
\end{tabular} 
\\
$^a$ The fraction in dust of iron is derived from the expression (\ref{Rho}) and represents a dust-to-metal ratio by number.  
In the $z=0.8597$ system towards QSO 0454+039
the measured [Zn/Fe]  ratio is lower
than the adopted   [Zn/Fe]$_\mathrm{g}$ for the  parameter sets E0 and E1;
in these cases we set $f_\mathrm{Fe}=0$ to avoid a negative, unphysical result. 
 For the same reason, the  error bars of $f_\mathrm{Fe}$ have been truncated
in a few cases.    
\\ 
$^b$ Logarithm of the observed column densities expressed in atoms cm$^{-2}$.
\\
$^c$ Parameter set Sa: [Zn/Fe]$_\mathrm{g}=0.0$ dex, $f_{\mathrm{Zn},{wd}}=0.59$; results independent of the adopted value of $\varepsilon_\mathrm{Zn}$.
\\
$^d$ Parameter set Sb: [Zn/Fe]$_\mathrm{g}=0.0$ dex, $f_{\mathrm{Zn},{wd}}=0.30$; results independent of the adopted value of $\varepsilon_\mathrm{Zn}$.
\\
$^e$ Parameter set E0: [Zn/Fe]$_\mathrm{g}=+0.1$ dex, $f_{\mathrm{Zn},{wd}}=0.59$;  $\varepsilon_\mathrm{Zn}=0$.
\\
$^f$ Parameter set E1: [Zn/Fe]$_\mathrm{g}=+0.1$ dex, $f_{\mathrm{Zn},{wd}}=0.59$;  $\varepsilon_\mathrm{Zn}=1$.
\\
$^g$ Sum of all components of the \ion{Fe}{ii} and \ion{Zn}{ii} column densities.
\\
--------------
\\
References.---
(1) Centuri\'on et al. 2003; 
(2) Centuri\'on et al. 2004 (work in preparation); 
(3) de la Varga et al. 2000;
(4) Dessauges-Zavadsky et al. 2003; (5) Ellison \& Lopez 2001; 
(6) Ge \& Bechtold 1997;
(7) Ledoux et al. 2002; (8) Ledoux et al. 2003; 
(9) Lopez et al. 1999; (10) Lopez et al. 2002;
(11) Meyer \& York 1992; (12) Meyer et al. 1995; 
(13) Molaro et al. 2000; 
(14) P\'eroux et al. 2002; (15) Petitjean et al. 2000; 
(16) Pettini et al. 1994; (17) Pettini et al. 1999; (18) Pettini et al. 2000;  
(19) Prochaska et al. 2001 (I database); (20) Prochaska et al. 2002 (IV database); (21) Prochaska et al. 2003 (ESI database);
(22) Prochaska \& Wolfe 1996; (23) Prochaska \& Wolfe 1997; (24) Prochaska \& Wolfe 1999.
}
\end{table*}
\end{center}

\subsection{The sample}

The data used in the present investigation were selected searching for DLA systems
with  published measurements of   \ion{Fe}{ii} and \ion{Zn}{ii} column densities.
All systems with $N(\ion{H}{i})$   in excess of the canonical threshold
$N(\ion{H}{i}) =10^{20.3}$ atoms cm$^{-2}$, or consistent with the threshold
 at $\simeq 1 \sigma$ level,
were included in the sample.  
Upper limits were excluded. 
The resulting sample, listed in Table 1,  includes 38 systems  in the redshift interval  
$0.6 \leq z \leq 3.4$. 
The references to the original measurements are given in  the table. 
Most of the measurements were obtained using high resolution spectrographs fed
by 8-10m class telescopes such as the Keck and the VLT.  
Care has been taken in checking that all the column densities are derived
with an updated and consistent set of oscillator strengths
(Bergeson \& Lawler 1993; Welty et al. 1999). 
%
  
   \begin{figure*}
   \centering
 \includegraphics[width=5.7cm,angle=0]{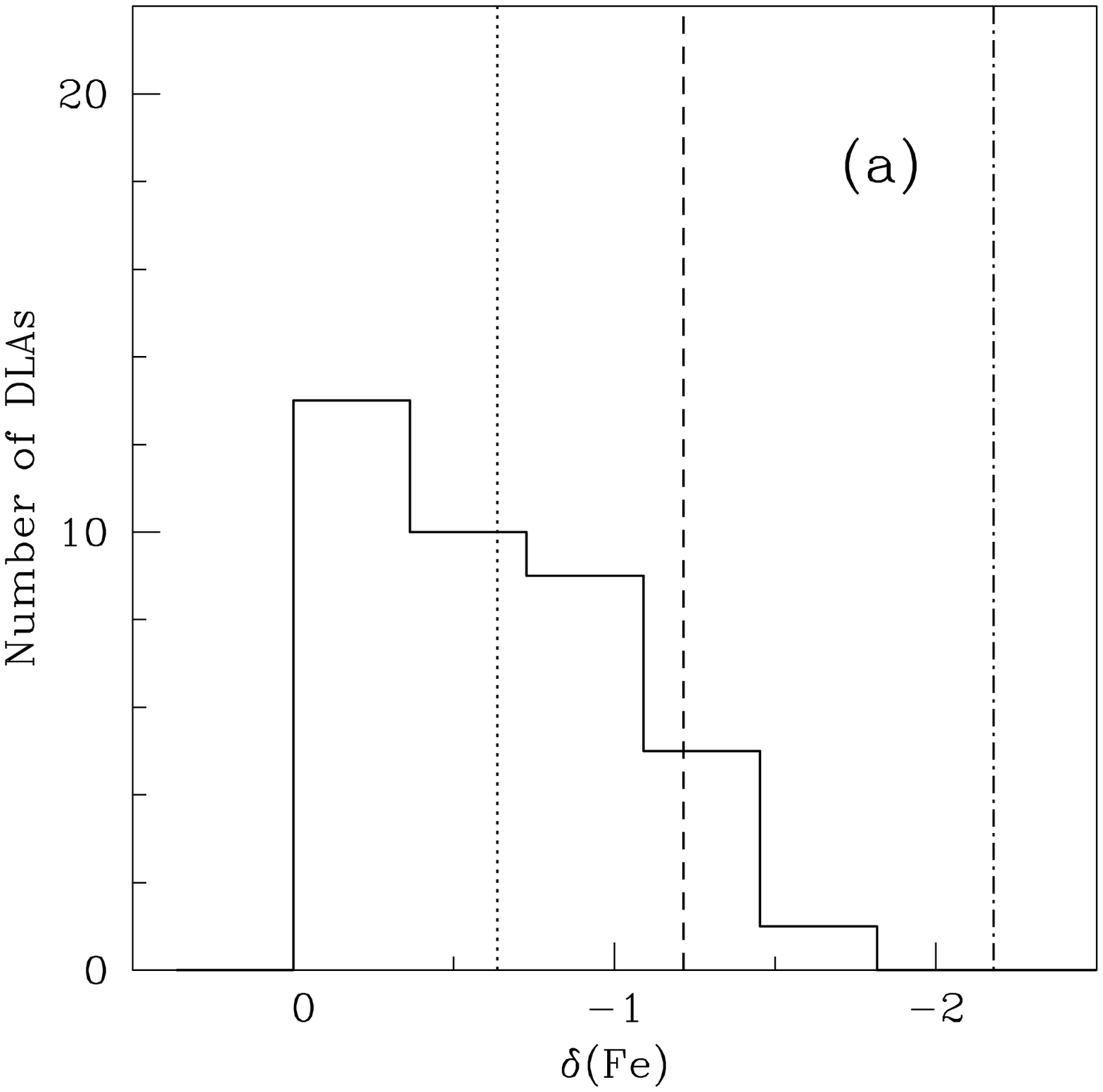} 
 \hspace{0.2cm}
  \includegraphics[width=5.7cm,angle=0]{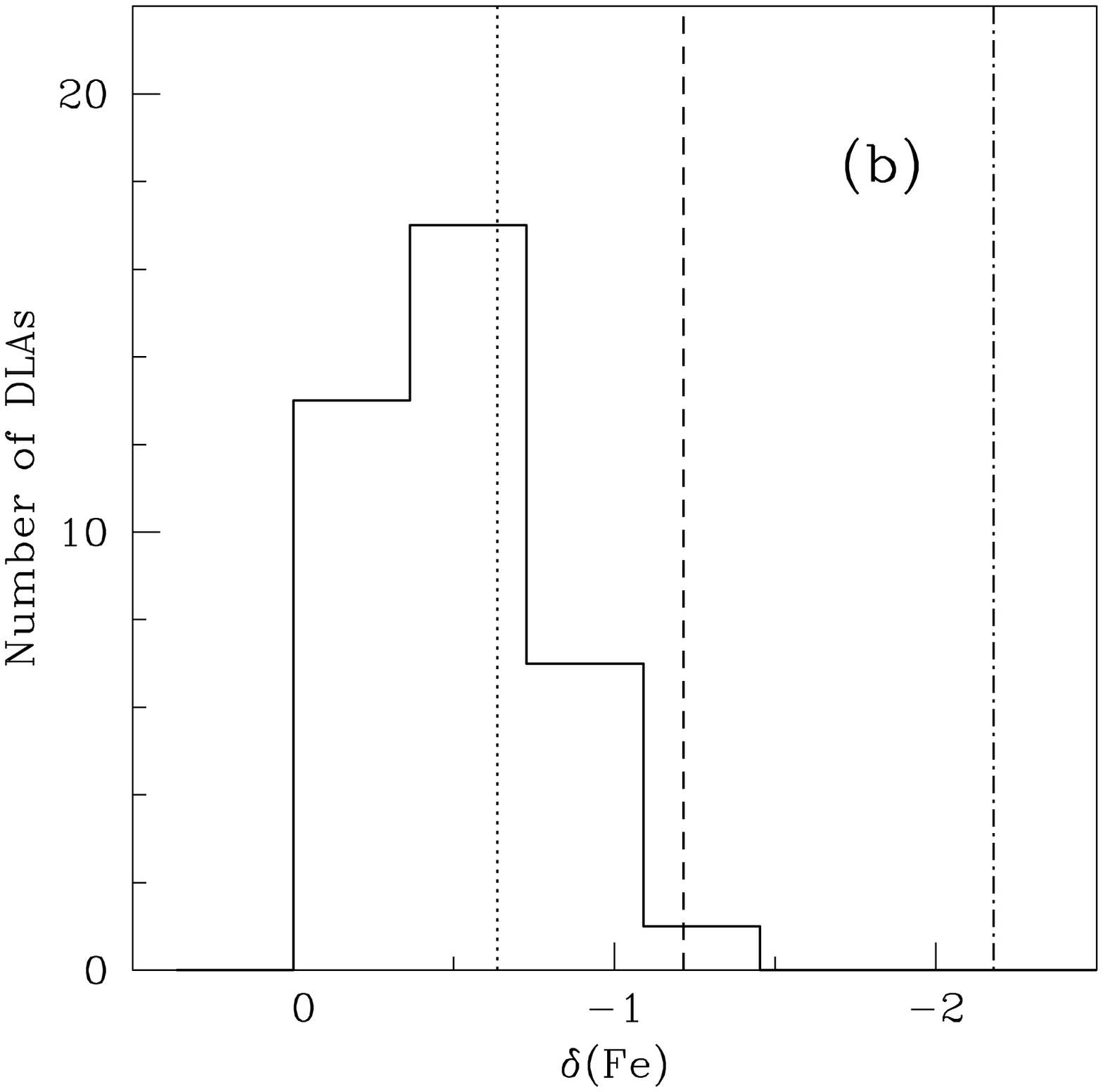} 
 \hspace{0.2cm}
  \includegraphics[width=5.7cm,angle=0]{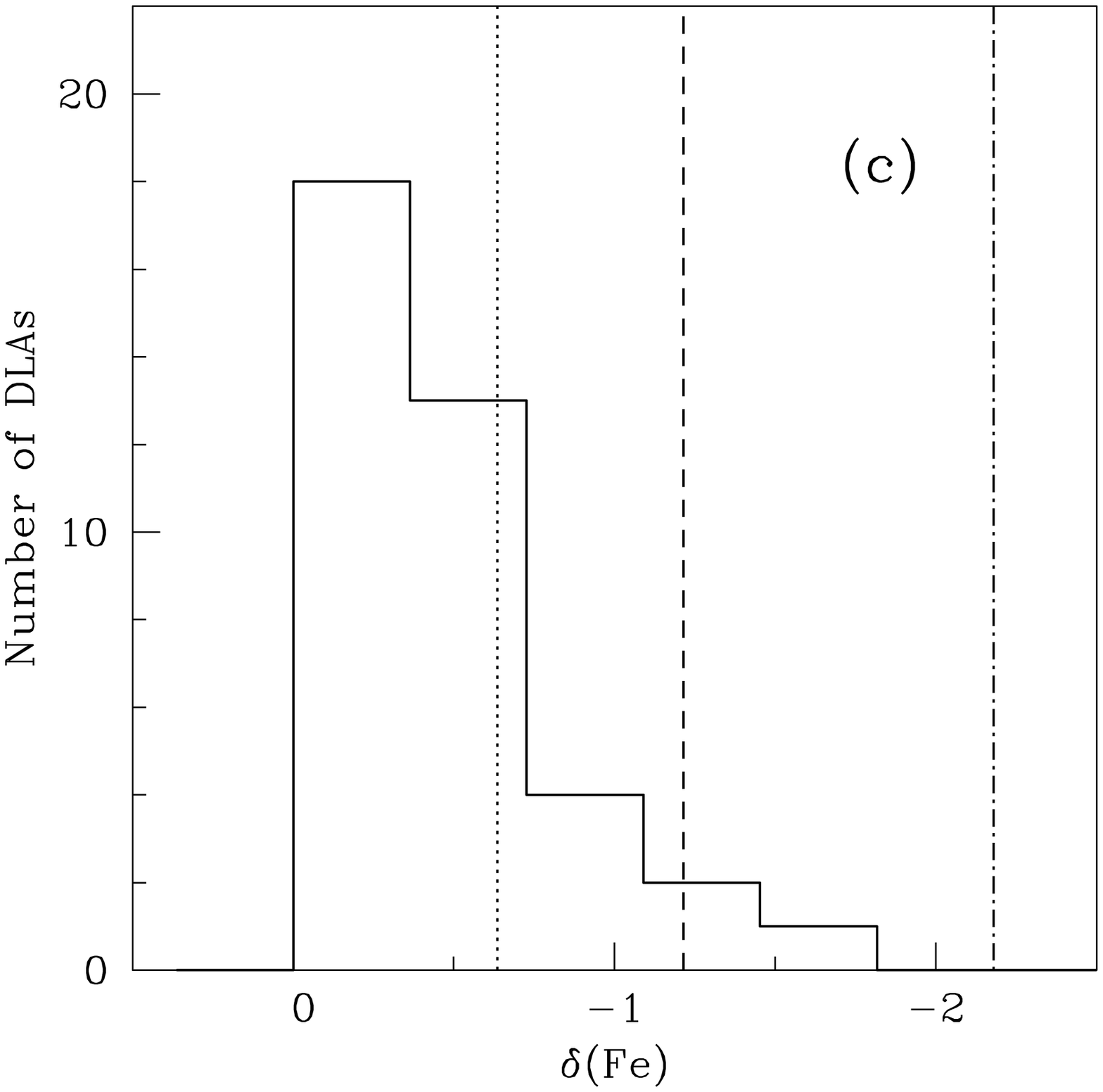} 
  \caption{Distribution of iron depletions for the DLA systems in Table 1.
      Panels {\bf a)}, {\bf b)} and {\bf c)}
      correspond to cases Sa, Sb and E1, respectively. Vertical lines:
         typical values of interstellar iron depletions in the
  Milky Way (dotted line: warm halo gas; dashed line:   warm disk gas;
  dotted-dashed line:   cold disk gas).
 }
              \label{DepHist}%
    \end{figure*}

\section{Results}

In the last 4 columns of Table 1 we list the
fraction in dust of iron in DLA systems $  f_\mathrm{Fe} 
=  \varrho \, f_{\mathrm{Fe},{wd}} $    
obtained from Eq. \ref{Rho} 
  for    different choices of
the input parameters $f_{\mathrm{Zn},{wd}}$, $\varepsilon_\mathrm{Zn}$
and [Zn/Fe]$_\mathrm{g}$, 
specified in the  notes to the table. 
The quoted errors have been derived by propagating the 
\ion{Fe}{ii} and \ion{Zn}{ii} column-density
errors.  
The results obtained for a solar [Zn/Fe]$_\mathrm{g}$ are labeled S;
those for an enhanced [Zn/Fe]$_\mathrm{g}(=+0.1)$ are labeled E. 
For the case S we list      
the results  obtained using $f_{\mathrm{Zn},{wd}}=0.59$
and $f_{\mathrm{Zn},{wd}}=0.30$, labeled Sa and Sb, respectively.
The differences are in most cases well below the quoted errors.
The results S do not depend on the adopted value of  $\varepsilon_\mathrm{Zn}$
(see Eq. \ref{Rho}). 
The results E do  depend on this
parameter 
and are labeled E0 and E1 to indicate the cases
$\varepsilon_\mathrm{Zn}=0$ and $=1$, respectively.
Comparing the columns E0 and E1 of Table 1 one can see that
the dependence on $\varepsilon_\mathrm{Zn}$ is indeed weak,
as anticipated above. %

\subsection{Frequency distribution of depletions}

As a first step of our analysis we consider the 
frequency distribution of iron depletions\footnote{
It is easy to show that  Eq. \ref{Depletion}  
is equivalent to the classical definition  
$
\delta_\mathrm{Fe}= 
\log [ N(\mathrm{Fe})_\mathrm{obs}/N(\mathrm{H})_\mathrm{obs} ]
- \log (\mathrm{Fe}/\mathrm{H})_{\sun}
$,
 when the intrinsic abundance of the medium is solar, as  in
   the Galactic ISM.
}, 
\begin{equation}
\label{Depletion}
\delta_\mathrm{Fe} = \log (1 - f_\mathrm{Fe}) ~,
\end{equation} 
 shown  in Fig. \ref{DepHist}.  
For the sake of comparison with      
Galactic interstellar studies we also plot 
the depletions 
representative
of   the Galactic  cold disk gas (dashed-dotted line), the warm disk gas
(dashed line), and the warm halo gas (dotted line), taken
from Savage \& Sembach (1996).

The frequency distributions are shown for   different choices of input parameters 
(cases Sa, Sb and E1).  
The main results common to all cases can be
summarized as follows.
The relative number of systems decreases with increasing depletion
$|\delta_\mathrm{Fe}|$.
A large fraction of systems have depletions {\em lower than
the lowest Galactic depletions}  (those of the warm halo gas; dotted line). 
The remaining DLAs have depletions similar to those of the warm
halo   or     disk gas; only for these cases it is correct to
conclude that  "depletions in DLA systems are 
 typical of warm Galactic gas", a claim often quoted in literature. 
Cold disk depletions are not found in the sample. 
These conclusions are also valid for the case E0, not shown
in the figure.  

   \begin{figure*}
   \centering
 \includegraphics[width=5.7cm,angle=0]{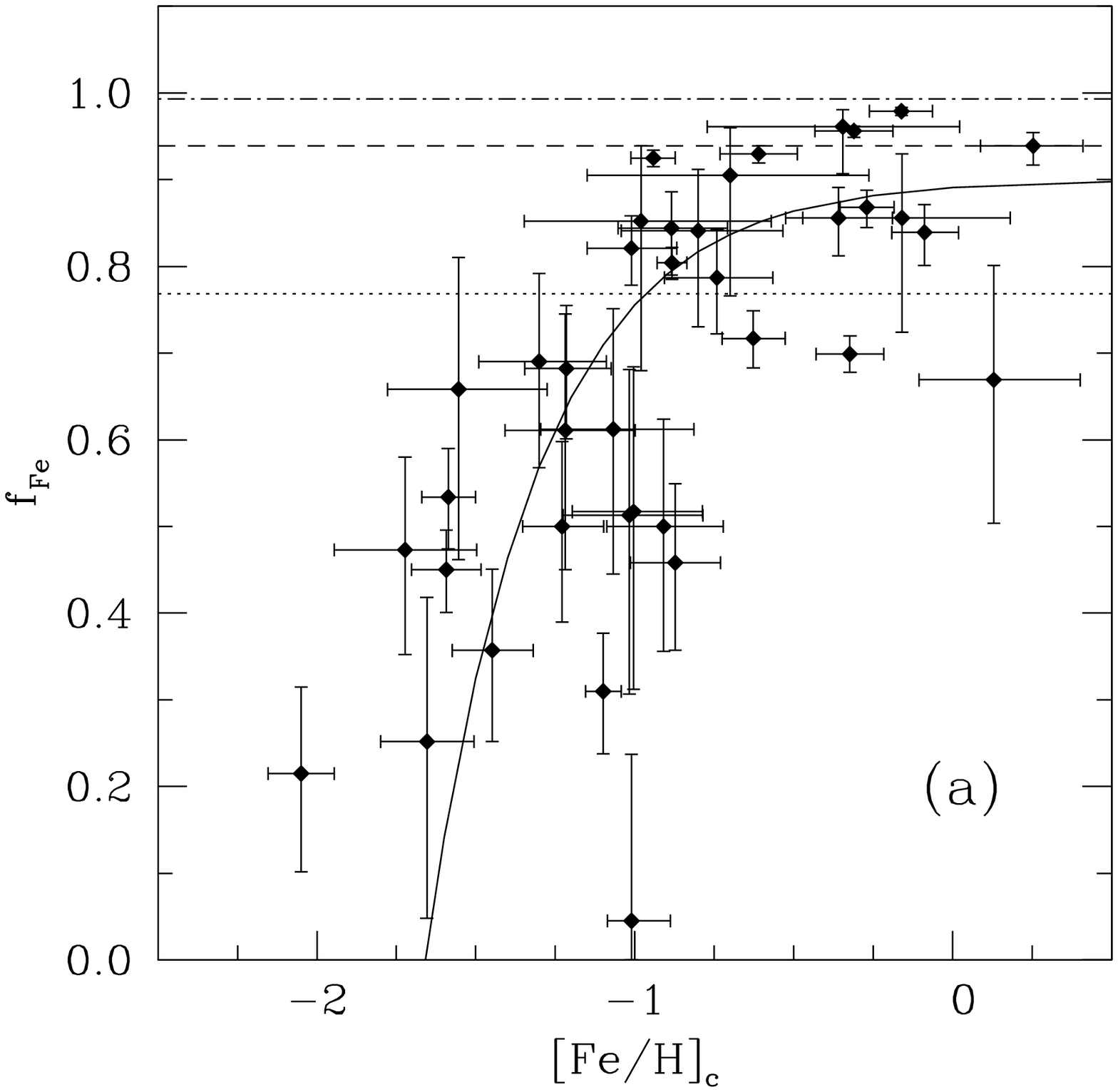} 
 \hspace{0.2cm}
  \includegraphics[width=5.7cm,angle=0]{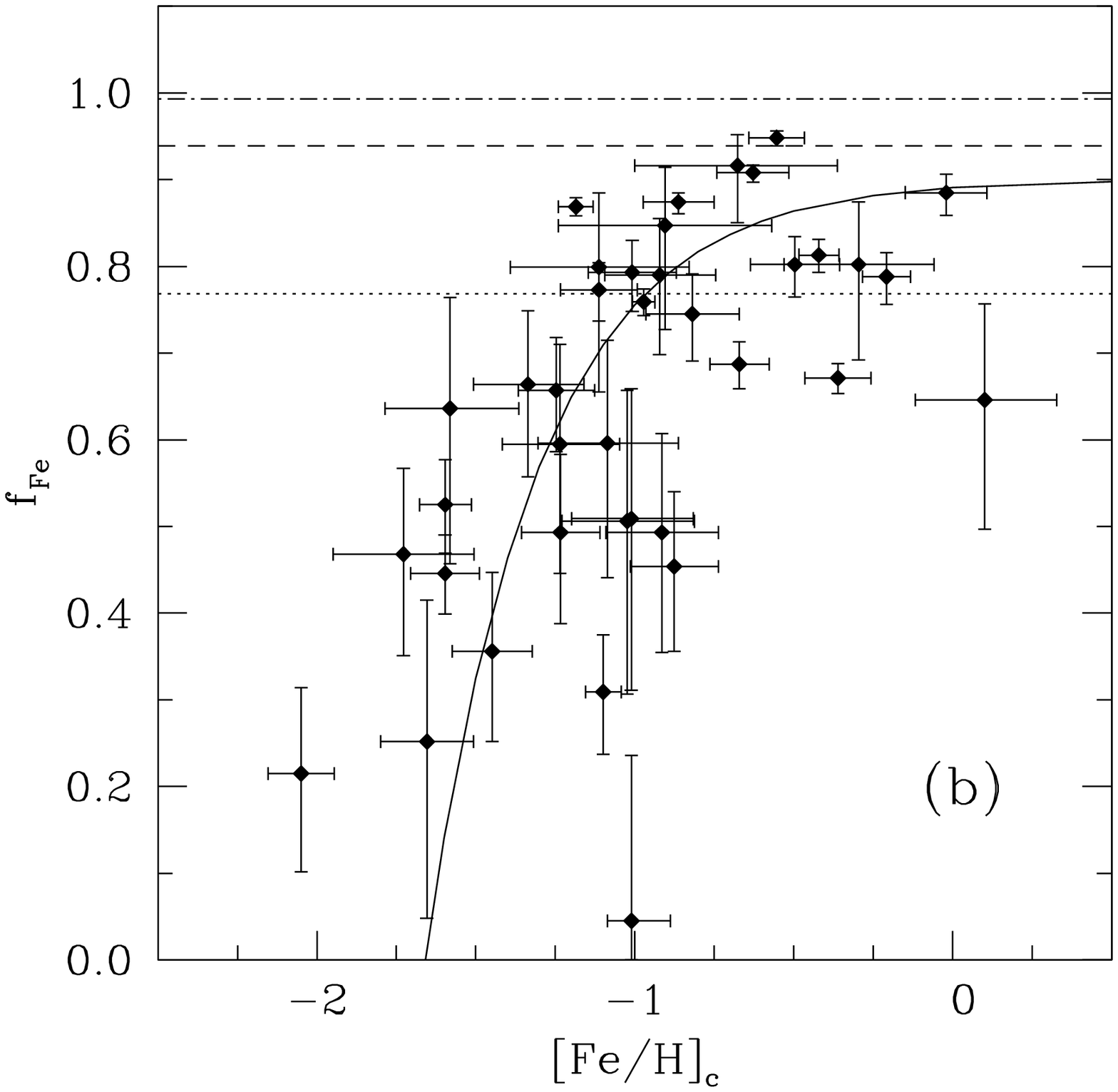}
 \hspace{0.2cm}
  \includegraphics[width=5.7cm,angle=0]{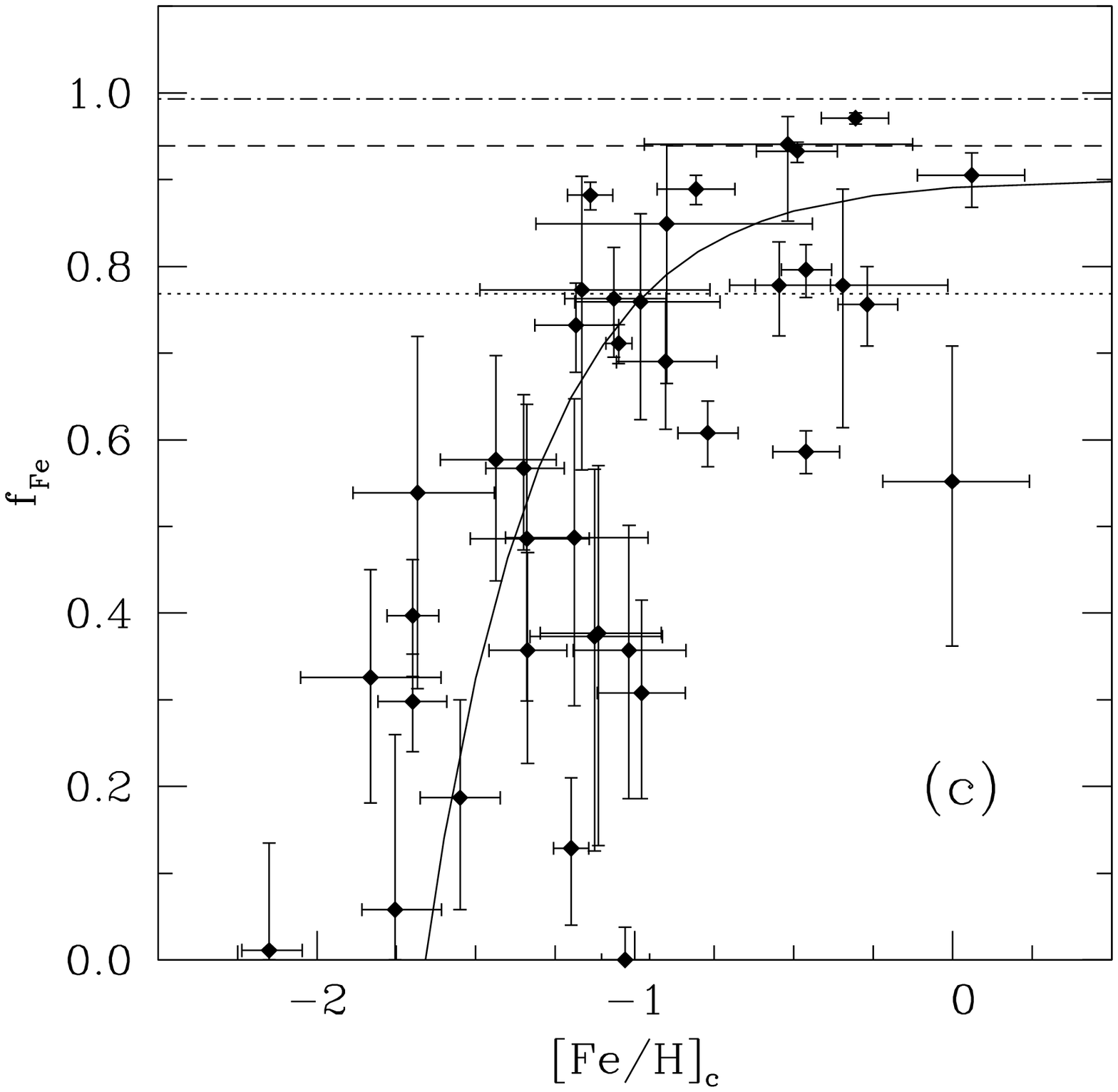}
  \caption{
  Iron  dust fraction 
  versus metallicity for the sample of DLA systems
  listed in Table 1. Panels {\bf a)}, {\bf b)} and {\bf c)}
  correspond to cases Sa, Sb and E1, respectively. 
  Horizontal lines:  representative values of   iron dust fractions in the
  Galactic ISM (legenda as in Fig. \ref{DepHist}). Solid curve:
  empirical law (\ref{f_law}) describing the evolution of the dust fraction with metallicity. }
              \label{DepMet}%
    \end{figure*}

\subsection{Dust fraction versus metallicity}

We next consider the relation between dust fraction,
$f_{\mathrm Fe}$, and metallicity,
which is shown in Fig. \ref{DepMet}
for  three different choices of input parameters
(cases Sa, Sb and E1).  A very similar relation is found
for the case E0, not shown
in the figure.  
The main result, common to all cases considered, is that
{\em the dust fraction increases with metallicity, with a flattening
at [Fe/H]$_{\mathrm c} \gsim -1$ dex
and a significant scatter}.  
 The data  follow  an empirical law of the type
\begin{equation}
f_\mathrm{Fe} \approx 
f_{\mathrm{Fe},\circ} -10^{\alpha(\mathrm{[Fe/H]}-\mathrm{[Fe/H]}_{\circ})}
\label{f_law}
\end{equation}
with 
$f_{\mathrm{Fe},\circ} \simeq 0.9$,
$\alpha \simeq -1.2$, and
$\mathrm{[Fe/H]}_{\circ} \simeq -1.7$
(solid curve in the figures). 
 The upper (lower) envelope of the data, shown in Fig. \ref{BiasTest}b, can be
 represented with the same law, with parameters
 $f_{\mathrm{Fe},\circ} \simeq 0.99$ $(0.67)$,
 $\alpha \simeq -1$ $(-4)$, and
$\mathrm{[Fe/H]}_{\circ} \simeq -1.7$ $(-2.15)$.
  
   The comparison with local ISM depletions  
reveals that the systems with  depletions 
lower than in the Galactic   halo gas
are mostly found at  [Fe/H]$_{\mathrm c} \lsim -1$ dex.
 
 We stress that the increase of  $f_{\mathrm{Fe}}$ with metallicity
is   confirmed even assuming  that [Zn/Fe]$_g$ evolves
from $\approx +0.2/+0.3$ dex  at [Fe/H] $\lsim -2$ dex, down to
$\simeq 0$ at solar metallicity, as claimed
by some authors (see Section 2.2.3). 
In fact, in this case 
we would obtain even less dust than in Fig.  2c
at [Fe/H] $\lsim -2$ dex, but the same amount
of dust as in Fig. 2a at solar metallicity. One can see that, as a consequence, the 
rise of $f_{\mathrm{Fe}}$ with metallicity would be even steeper.  
We now discuss the stability of the result in light of possible 
observational bias.

\subsection{Selection bias}
 
 Several types of observational bias are known to affect the 
measurements of the  $N(\ion{H}{i})$, $N(\ion{Fe}{ii})$, and $N(\ion{Zn}{ii})$  
column densities.   %
For instance, the detection limit\footnote
{
The detection limits are estimated at 3$\sigma$ level for
a signal-to-noise ratio $S/N \simeq 30$  and
resolving power $R = \lambda/(\Delta \lambda) \simeq 5 \times 10^4$
at $\lambda \simeq 500$ nm. 
Oscillator strengths
for the \ion{Zn}{ii} 202.6136 nm
and the \ion{Fe}{ii} 160.8451 nm lines are taken from
Bergeson \& Lawler (1993) and from
Welty et al. (1999), respectively. 
}
 of the strongest line of the \ion{Zn}{ii} resonance doublet
prevents the measurement of column densities   
$N(\ion{Zn}{ii}) \la 10^{11.75}$ atoms cm$^{-2}$.
The detection limit of \ion{Fe}{ii} 
depends on the wavelength coverage of the spectra,
since  a variety of strong \ion{Fe}{ii} transitions are present
at different wavelengths. 
Assuming that we do not cover the stronger transitions
at longer rest wavelengths, we obtain a conservative limit
from  the detection limit of the 160.8 nm line,
$N(\ion{Fe}{ii}) \la 10^{12.88}$ atoms cm$^{-2}$.
Also  \ion{H}{i}   measurements are affected by selection bias.
The range of $N(\ion{H}{i})$ 
is limited  on one side by the definition threshold in DLA systems
$N(\ion{H}{i}) \geq 10^{20.3}$ atoms cm$^{-2}$
and, on the other side, by the fact
that values above $N(\ion{H}{i}) > 10^{20.85}$ atoms cm$^{-2}$
have never been observed, even if they are detectable. 
Finally, 
the  obscuration of the background QSO by dust extinction in the DLA
system could be responsible for the lack of systems
with high metal column densities, namely those with
 $N(\ion{Zn}{ii}) \ga 10^{13.1}$ atoms cm$^{-2}$ (Boiss\'e et al. 1998),
 which are also detectable. 
 
   \begin{figure}
   \centering 
\includegraphics[width=5.7cm,angle=0]{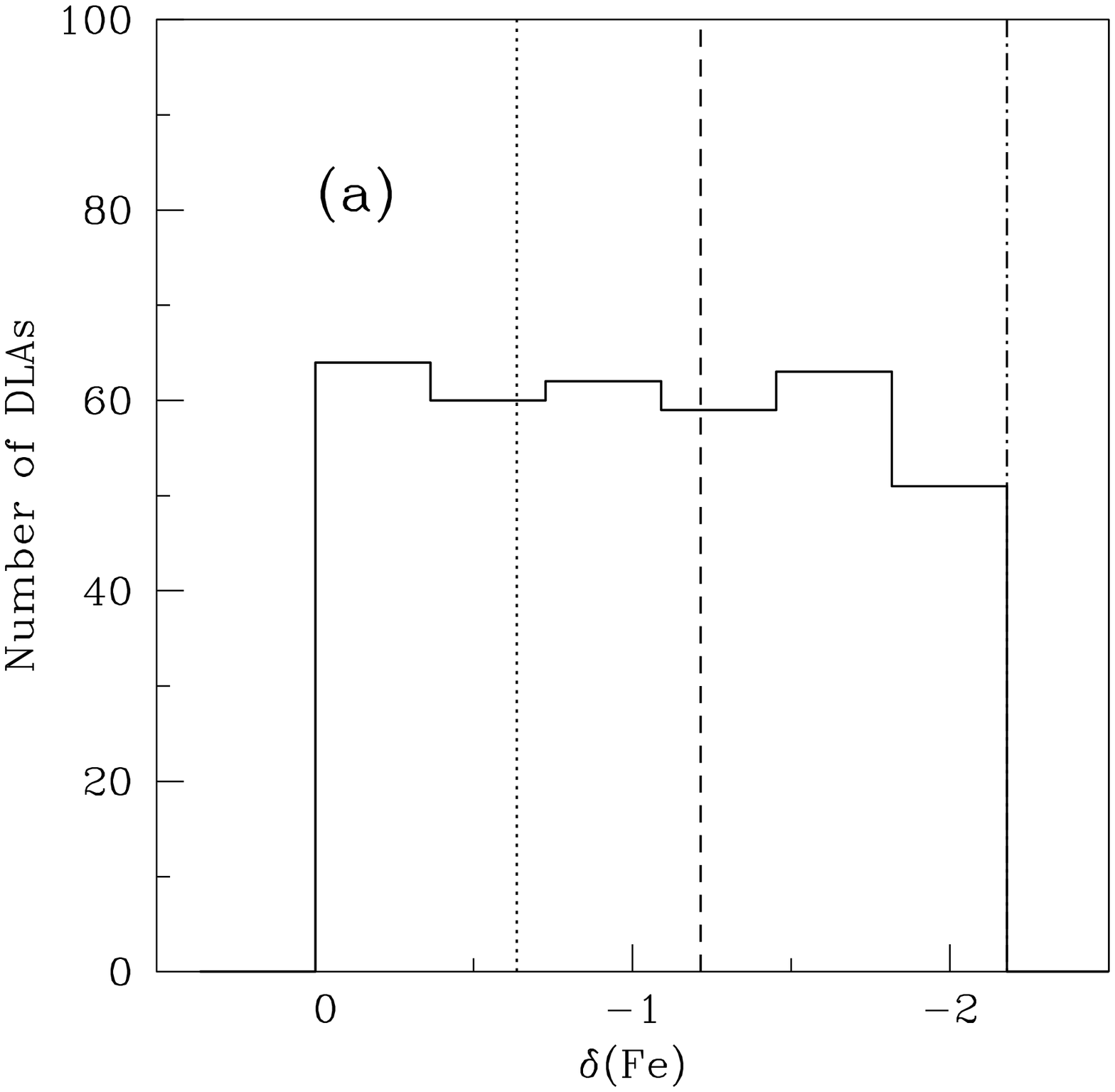} 
 \hspace{0.5cm} 
\includegraphics[width=5.7cm,angle=0]{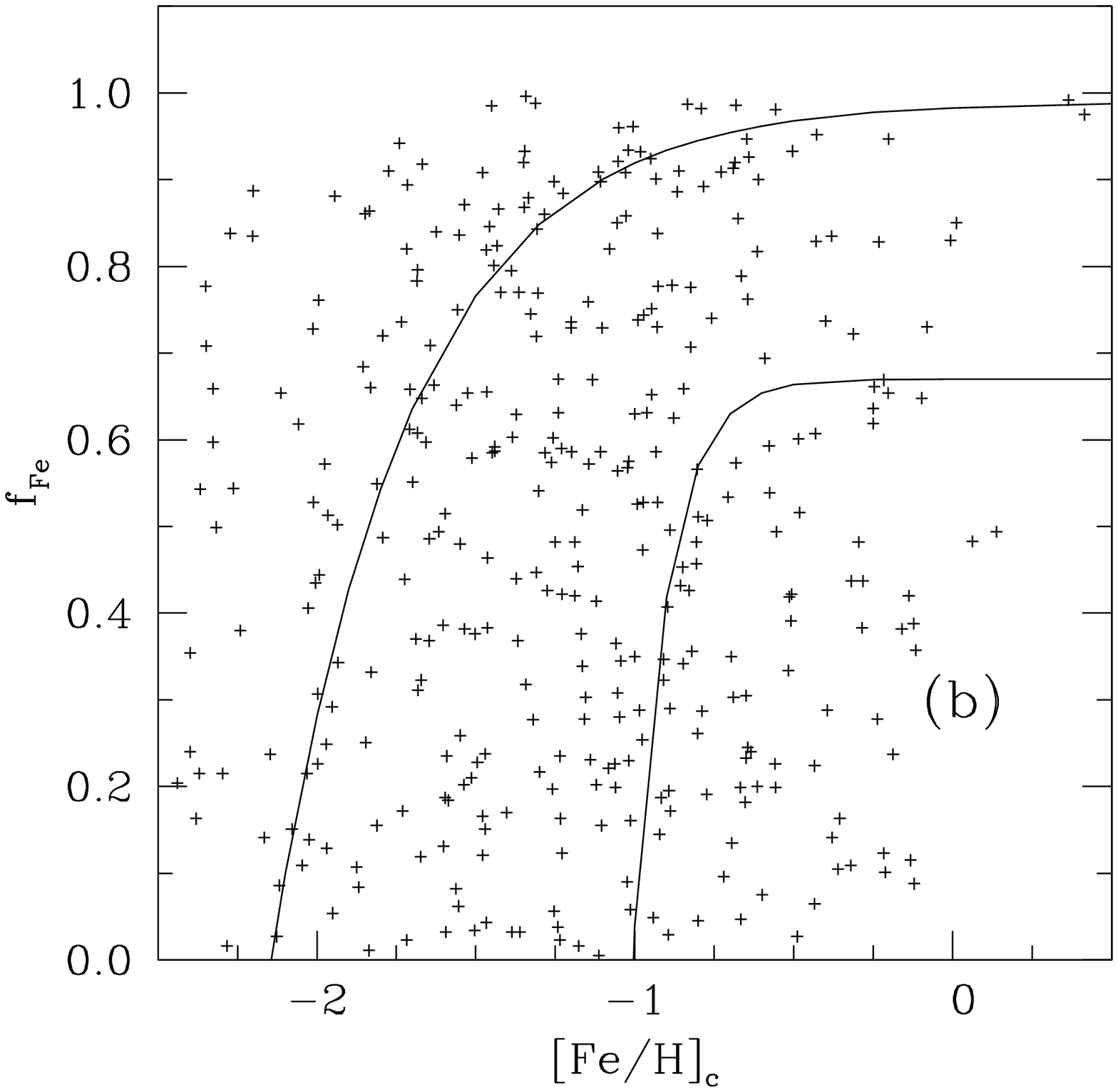}  
\caption{ Test of selection effects performed on 
  a mock sample of $n=1000$  DLA systems with   
    \ion{H}{i}, \ion{Fe}{ii} and \ion{Zn}{ii} column densities
  rejected  as explained in Section 3.3. 
   {\bf a)} 
  Frequency distribution of iron depletions resulting (after rejection)
  for  
  an adopted   flat distribution of $\delta$(Fe).  
    {\bf b)} Dust fraction versus metallicity resulting
  for an adopted  
  flat distribution of $f_\mathrm{Fe}$;
  the solid curves represent the envelopes of the real data in
  Fig. \ref{DepMet}. 
    The depletions of the mock sample  have been modeled with 
  the set of parameters  Sa (Table 1);
 similar results are obtained for the other cases. 
       }
              \label{BiasTest}%
    \end{figure}

 In order to assess the effect of these observational bias
 on the present study  
we  built a mock sample of  $n$ DLA systems with a flat 
distribution of metallicities and depletions (or dust fractions).
 We then assigned to each system a random-generated  \ion{H}{i}
column density in the allowed range
 $20.3 \leq \log N(\ion{H}{i}) \leq 20.85$
 and computed  
 $N(\ion{Fe}{ii})$ and $N(\ion{Zn}{ii})$  
   for each pair of 
([Fe/H],$f_\mathrm{Fe}$) values. 
Finally, we rejected cases 
with $N(\ion{Zn}{ii})$ and $N(\ion{Fe}{ii})$
outside the   limits discussed above.
  At this point, we applied to the   biased sample
 the same method for the derivation of dust fractions
 and metallicities described
 in Section 2.1 and the same  analysis
 presented in Sections 3.1 and 3.2. 
In this way, we   tested the potential effects 
 of  the column-density bias 
 and, at the same time,  of the application of 
 the method itself.   The results  are shown in Fig. \ref{BiasTest}
 for $n=1000$ systems (before rejection). 
   
 The frequency distribution of depletions of the biased mock
 sample is shown in  the   panel (a) of the figure. 
 One can see that, in spite of the large fraction of data points
 rejected  (65\%), the distribution  is still approximately flat  
 in the range $-2.2 \leq \delta(\mathrm{Fe}) \leq 0 $,
 as defined by construction. 
This indicates that the overall distribution of depletions 
is not significantly affected by bias and, therefore,
 the steepness    
of the real distribution  in Fig. \ref{DepHist}  is genuine.
We conclude that
{\em the   deficiency  of systems with high levels of dust depletion 
seems to be an intrinsic property of DLA systems rather than
a consequence of selection effects}.  

 The plot of dust fraction versus
 metallicity, for the   mock sample
 built to uniformly cover  the range $0 \leq f_\mathrm{Fe} \leq 1$,
 is shown in Fig. \ref{BiasTest}b. 
 The  non-rejected data of the mock sample
(crosses in the figure)
are able to populate regions
outside the boundaries in which the real data are confined 
(solid curves). 
This indicates that the selection bias
is not   responsible for
the lack of real data outside these boundaries.
In addition, the rather homogeneous
distribution of the data in  Fig. \ref{BiasTest}b,
derived from the application of our method
to the mock column densities, 
indicates that the method itself   does not 
induce  artificial trends 
between $f_\mathrm{Fe}$ and [Fe/H].
 We conclude that {\em the rise of  
 dust fraction with increasing metallicity 
 is a genuine property of DLA systems}. 
  
Clearly, a  better understanding of the dust obscuration effect
is required in order to take into account the role of selection bias
in a more realistic way.   
In any case, we expect the extinction of the QSO to be more and
more effective with increasing metallicity of the  
intervening absorbers.
Therefore,
the fact that  DLA systems with high dust content
($f_\mathrm{Fe} \gsim 0.8$)  {\em are} detected  at high
metallicity demonstrates that the obscuration bias cannot be responsible
for   the lack of such systems  at {\em low} metallicity
(compare the data points
with  [Fe/H] $\lsim -1.5$ dex and $\gsim -1.5$ dex in Fig. \ref{DepMet}).
On the other hand, the number of absorbers with depletions
typical of cold interstellar gas {\em and} high metallicity {\em can} be
affected by dust obscuration.
Therefore the deficit of cold gas  depletions 
could be less severe than shown  in Fig. \ref{DepHist},
but only for the systems of high metallicity.

     \begin{figure*}
   \centering
 \includegraphics[width=5.7cm,angle=0]{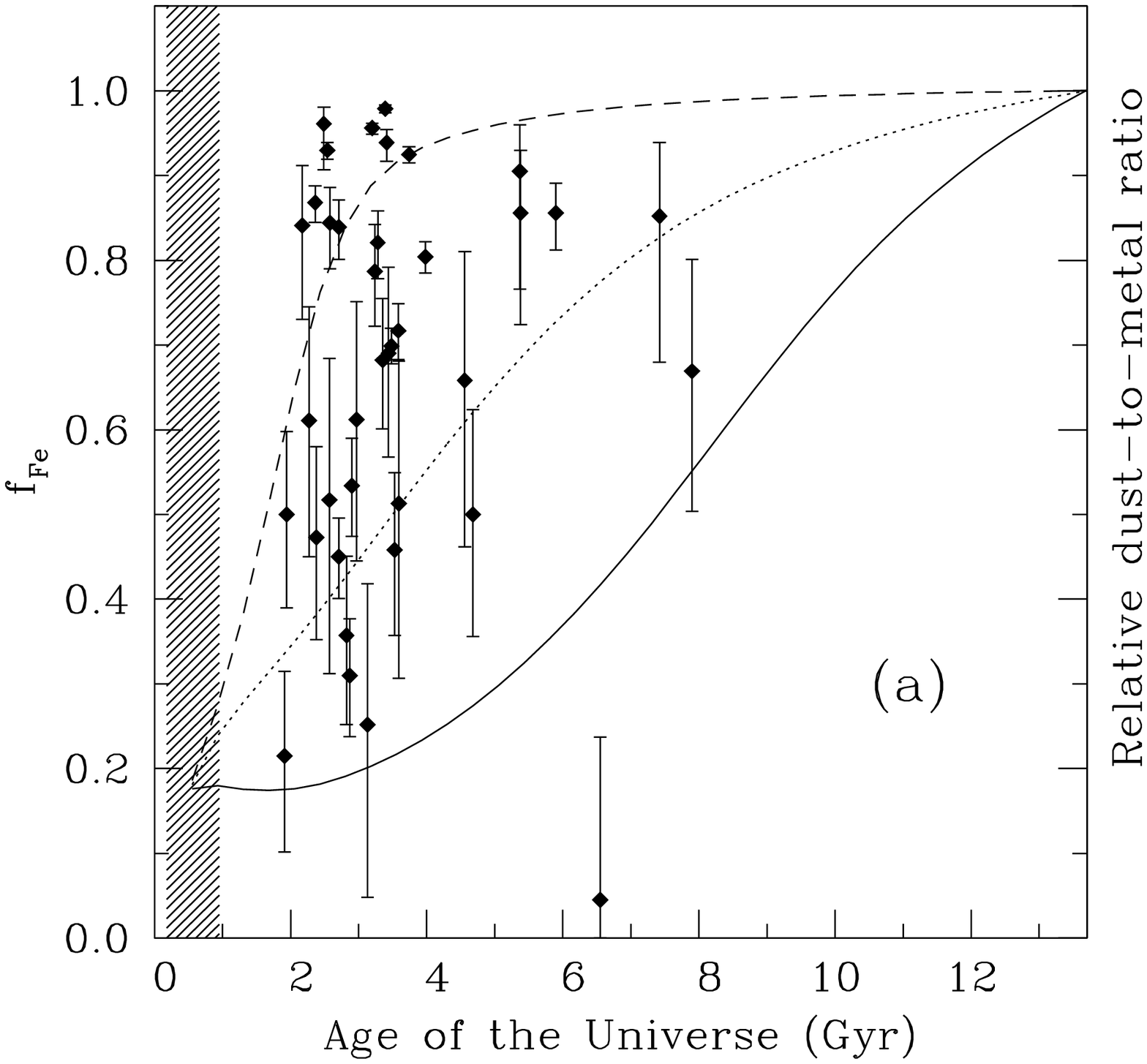} 
 \hspace{1.2cm}
  \includegraphics[width=5.7cm,angle=0]{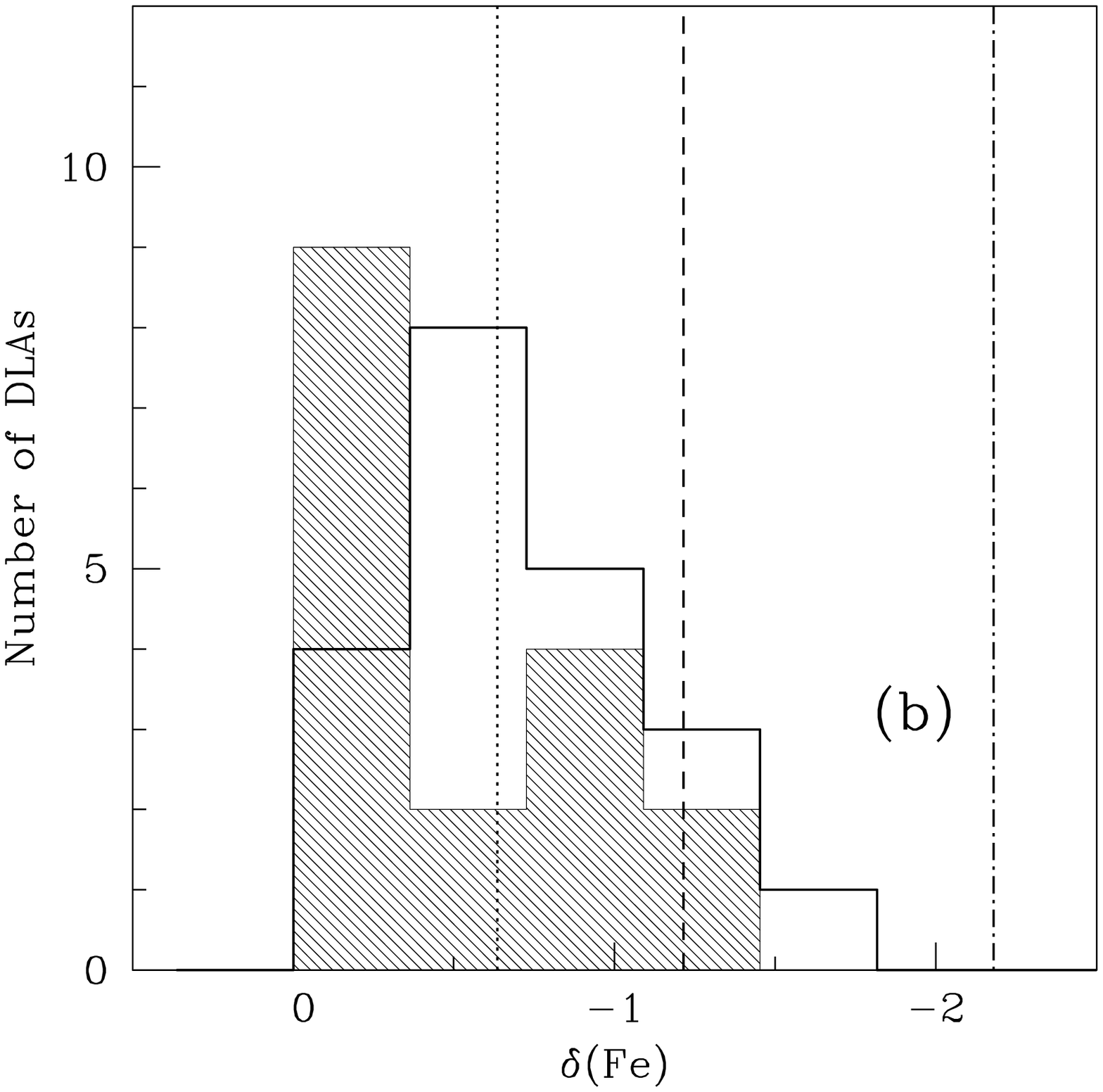} 
  \caption{
  Time evolution of the   dust content   in DLA systems.
 {\bf a)}    Diamonds: $f_\mathrm{Fe}$ versus age of the universe;
  dashed area:  reionization epoch  $ 6 \leq z \leq 20 $
  (Fan et al. 2003; Spergel et al. 2003); solid, dotted and dashed curves:
Inoue's (2003) model predictions of the  evolution of the relative dust-to-metal ratio 
for different sets of input parameters  (see text for explanations);
the curves have been normalized to the present-day value and scaled assuming 
a formation redshift  $z_f=20$.    
  {\bf b)} Frequency distribution of iron depletions
  in two
  bins of look-back time (shaded area: $t_l \ge 10.5$ Gyr;
  empty histogram: $t_l < 10.5$ Gyr). 
  The fraction in dust are computed for the set of parameters Sa.   }
              \label{DepTime}%
    \end{figure*}

\subsection{Time evolution}

Finally we study 
the time evolution of the dust fraction after
converting the redshift of each absorber
 to the look-back time 
or age of the Universe\footnote{
We calculated these conversions
 for a flat universe with H$_0=71$ km s$^{-1}$ Mpc$^{-1}$
  and $\Omega_m=0.27$ (Spergel et al. 2003).   }.    
The advantage of this approach is that we compare  $f_\mathrm{Fe}$
with a completely independent observable, while in
the study versus metallicity we compare quantities
derived from the  measurements of the same column densities. 
However,  
 the interpretation of the  time evolution  is complicated by the fact that  
 DLA galaxies may start their chemical evolution at
different cosmic times. 
 Inspection
of the plot of  $f_\mathrm{Fe}$ versus age of the
Universe, shown in Fig. \ref{DepTime}a, 
does not reveal a clear trend, even though the data are consistent
with a sudden   rise
of the dust-to-metal ratio between $\approx 2$ and $4$ Gyr after the Big Bang.
 Some evidence of evolution is found
by binning all the systems seen before and after the median value
of look-back time of the present sample, $t_l \simeq 10.5$ Gyr, 
and comparing the frequency
distribution of their depletions 
(Fig. \ref{DepTime}b).
One can see that  the two distributions are slightly shifted,
the DLAs seen at the earliest age of the universe (shaded histogram)
having lower
depletions $|\delta(\mathrm{Fe})|$ than those observed in more recent times
(empty hystogram). 
Again, these results hold valid for all the
possible choices of input parameters (case Sa, shown in the figure, and
 cases Sb, E0 and E, not shown).   
 
\section{Discussion}

\begin{figure*}
  \centering 
  \includegraphics[width=5.7cm,angle=0]{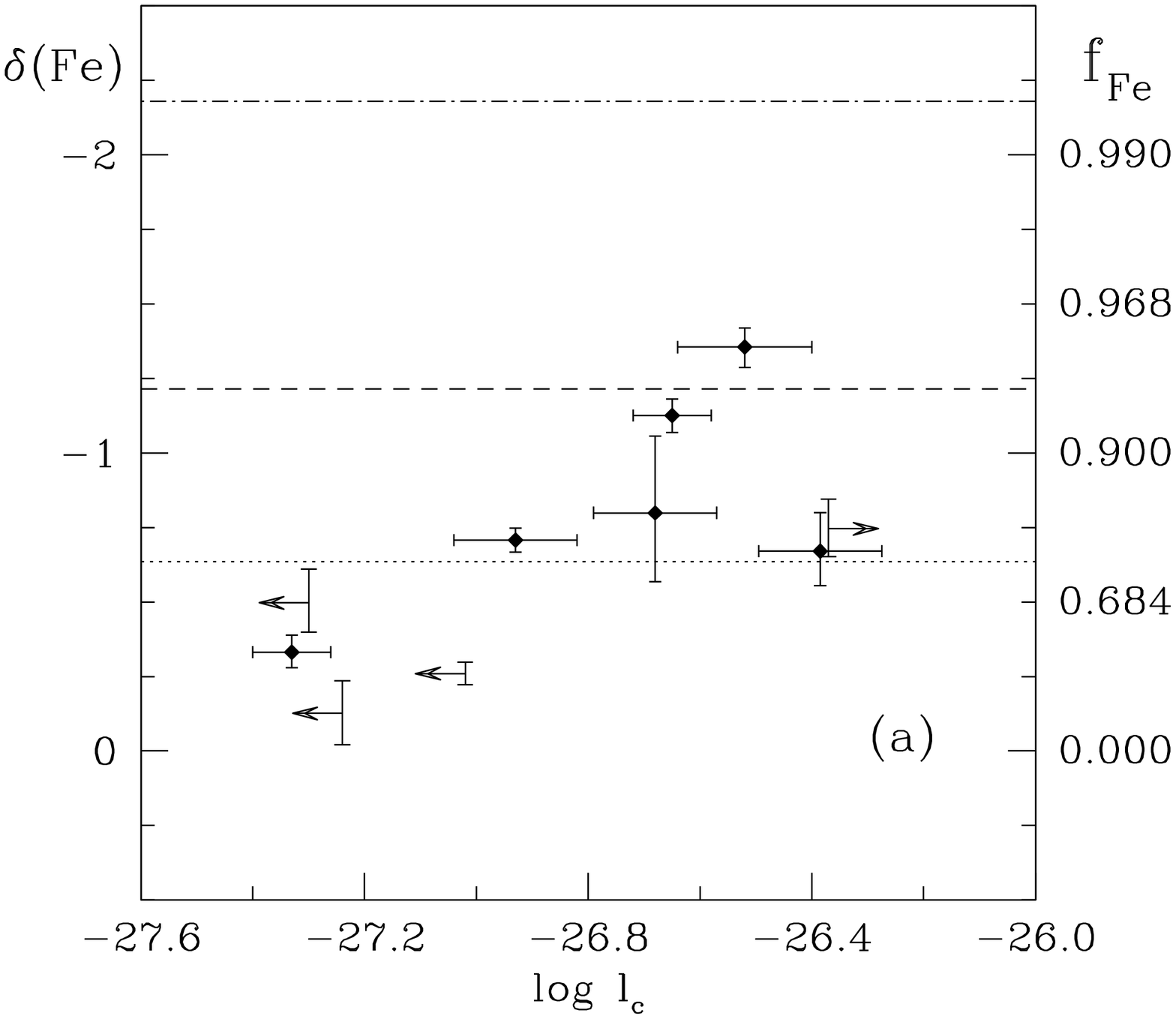}  
  \hspace{0.5cm}
  \includegraphics[width=5.7cm,angle=0]{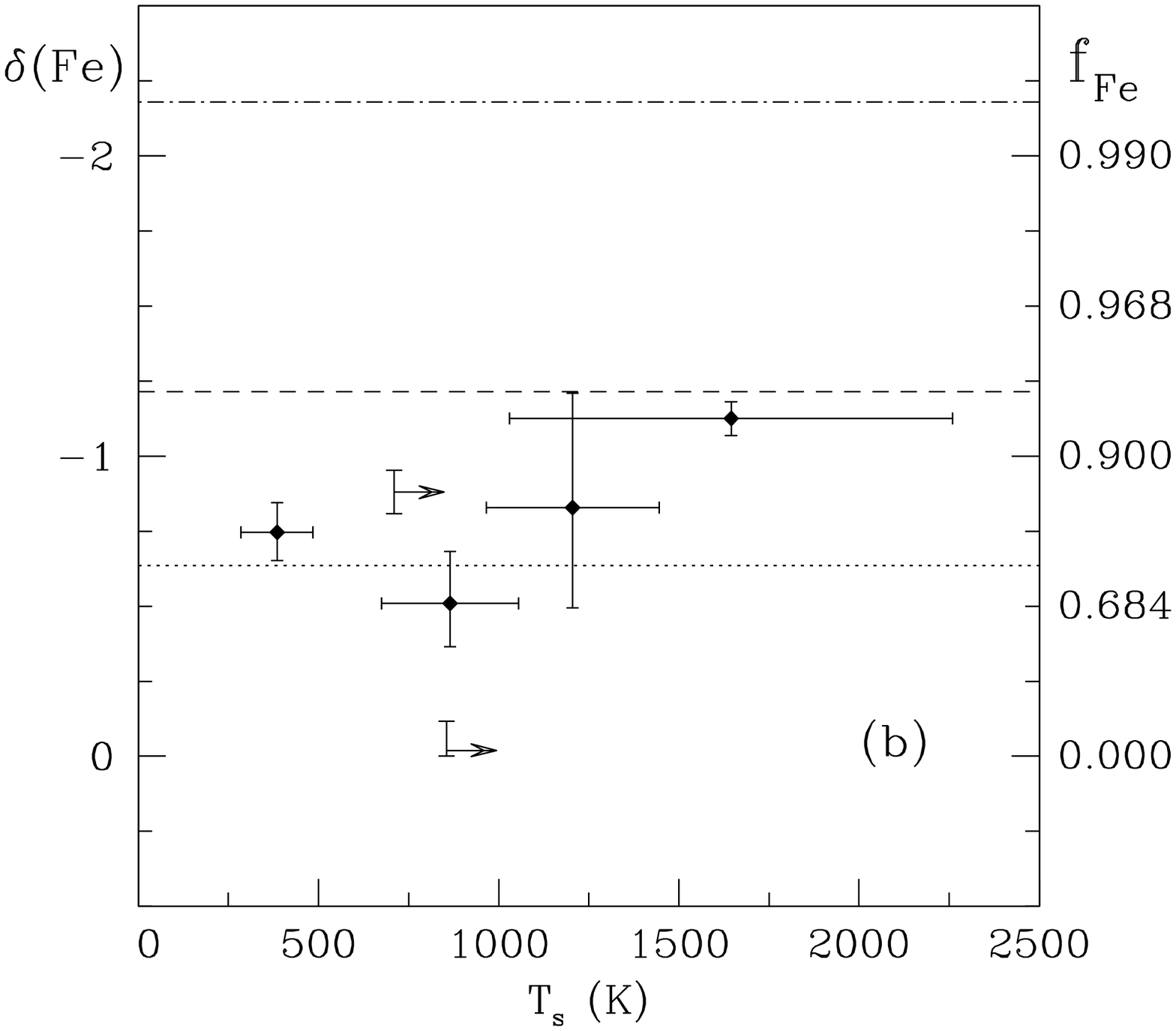}  
  \hspace{0.5cm}
  \includegraphics[width=5.7cm,angle=0]{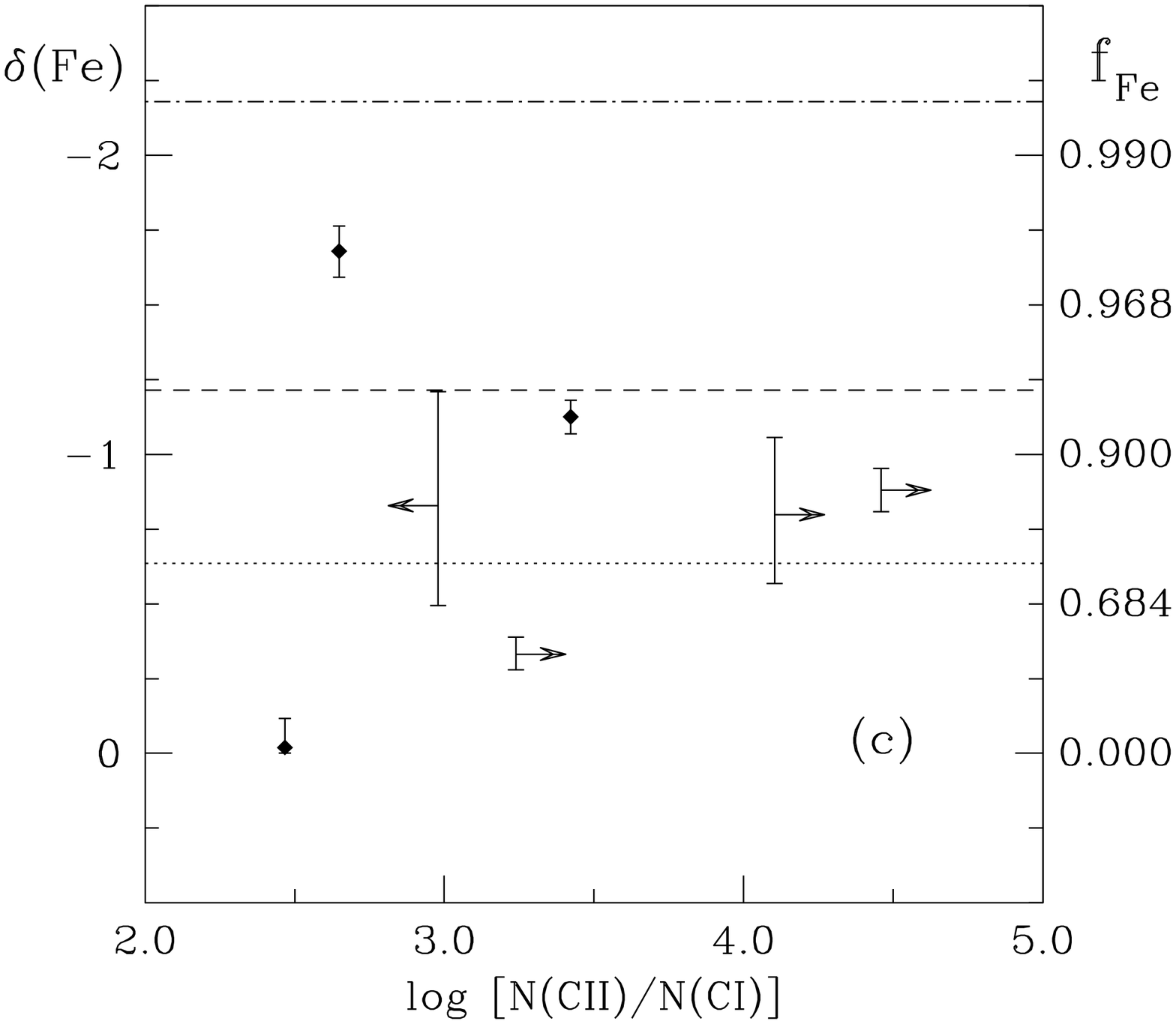}  
 \hspace{0.5cm}
 \includegraphics[width=5.7cm,angle=0]{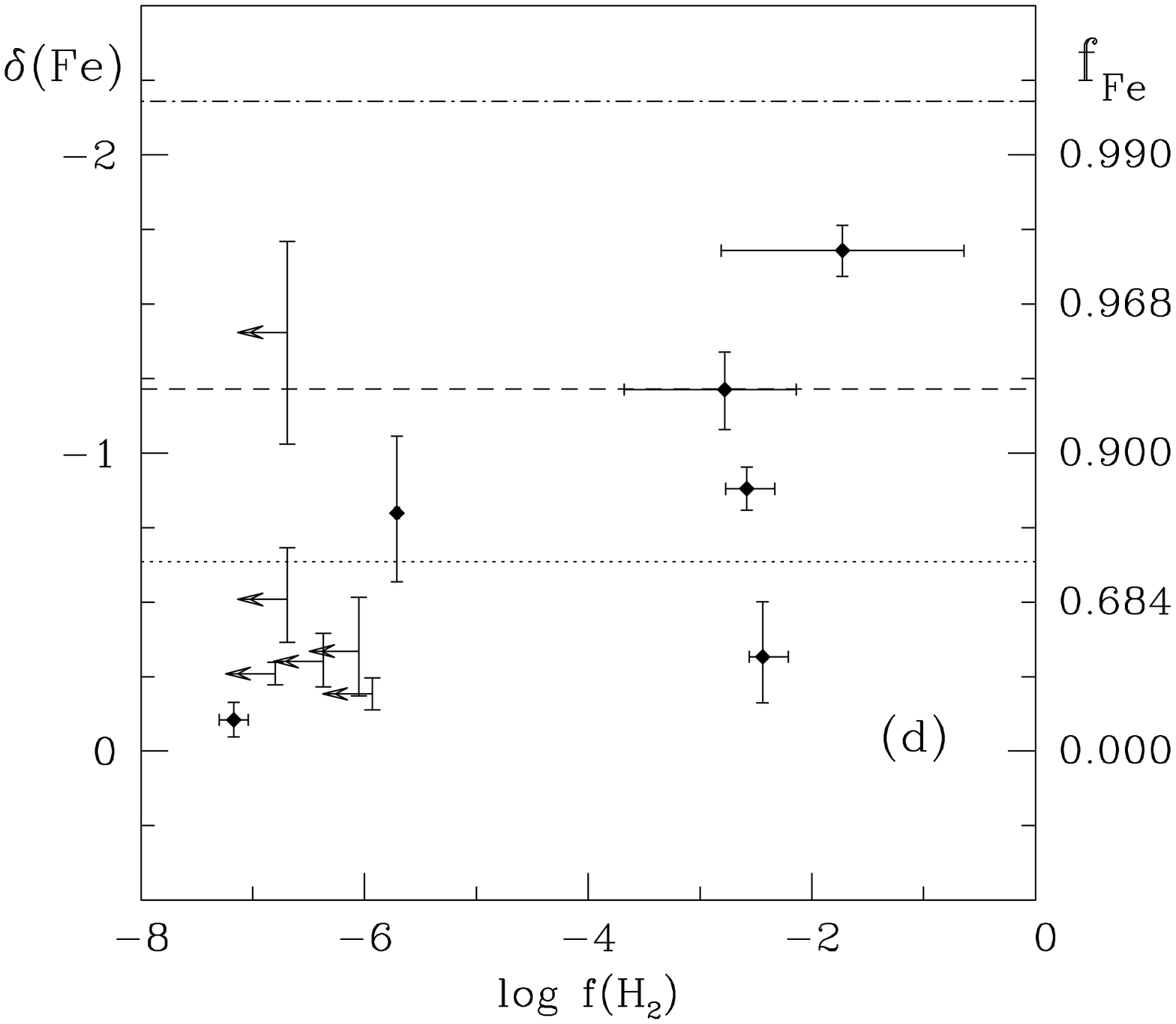}  
  \caption{ Iron depletions/dust fractions versus some physical parameters measured
  in DLA systems.
 {\bf a)} Cooling rate per H atom, $l_c$ (Wolfe et al. 2003a).
 {\bf b)} Spin temperature, $T_s$ (Kanekar \& Chengalur 2003).
 {\bf c)} $N(\ion{C}{ii})/N(\ion{C}{i})$ ionization ratio (Liszt 2002).
 {\bf d)} Molecular fraction $f(\mathrm{H}_2)$ (Ledoux et al. 2003).
 Depletions are computed in case Sa. Horizontal lines as in Fig. \ref{DepHist}. 
 }
              \label{TestPhys}%
    \end{figure*}

We now discuss the empirical results presented 
in the previous section
assuming that they are substantially unaffected by selection bias.
%
We first try to understand whether the observed trends 
are mainly governed by variations of the   
physical properties  or by chemical evolution.  
An example of the difficulty of disentangling these two effects
is given in Fig.  \ref{TestPhys}a, where
we plot the iron depletions versus $l_c \propto N(\ion{C}{ii^*})/N(\ion{H}{i})$,    
an indicator of the cooling rate of the gas  
(Wolfe et al. 2003a). Apparently, $|\delta(\mathrm{Fe})|$ seems
to increase with $\log l_c$ and 
this trend could be taken as evidence 
for a correlation of the depletions with 
some physical property  of the gas.
However, $l_c$ is by definition  proportional
to the metallicity  and the trend in Fig.  \ref{TestPhys}a may simply
mirror the rise of $f_\mathrm{Fe}$ 
with metallicity
shown in Fig. \ref{DepMet}. 
We now consider other observational evidence
that might help in disentangling the role  of physical processes.

\subsection{ The dependence on physical parameters}

In the Galactic ISM, each interstellar phase is characterized
by a particular level of depletion. 
Cold, high-density gas, where the dust has higher accretion probability
and is more shielded by destruction processes,
is characterized by high depletions;
warm, low-density gas is instead known to have
 low  depletions (e.g. Savage \& Sembach 1996).
This link  
  between the physical state  and level
of depletion might be at work also in the ISM of other galaxies,
including 
DLA systems. 
If so,  low values of depletions would map
regions of warm gas, while higher depletions colder regions.
According to this interpretation, the frequency distribution in
Fig. \ref{DepHist} would indicate that in DLA systems warm regions
are more frequent (have a higher cross-section)
than cold regions. 
The evolutionary trends with metallicity (Fig. \ref{DepMet}) and cosmic time
(Fig. \ref{DepTime}) would indicate that cold regions are extremely rare in the
early stages of evolution, but eventually they become more frequent. 

Some evidence in favour of this interpretation comes from 
measurements of the spin temperature,
$T_s$ (K), derived 21-cm absorption observations
(Carilli et al. 1996, Chengalur \& Kanekar 2000, Kanekar \& Chengalur 2003)
and from 
studies of the $N(\ion{C}{ii})/N(\ion{C}{i})$ ratio
(Liszt 2002),
which suggest that the warm phase
is generally dominant in DLA systems, especially at high redshift
(see however Wolfe et al. 2003b). 
Nevertheless,
a   detailed study of the results on a system-by-system basis
does not support a direct connection
between depletions and physical parameters in DLAs. 
In fact, 
if the observed trends of depletions were governed by 
variations of the physical properties, we would expect to find 
a decrease of $|\delta(\mathrm{Fe})|$ with increasing   $T_s$  
or 
$N(\ion{C}{ii})/N(\ion{C}{i})$, i.e. with  
increasing importance of the warm component. 
The   comparison of depletions versus $T_s$ for the systems
in common with the compilation
by Kanekar \& Chengalur (2003) 
does not support this expectation
(Fig. \ref{TestPhys}b). Also the comparison with the
 $N(\ion{C}{ii})/N(\ion{C}{i})$ determinations  by Liszt (2002) 
does not reveal the expected trend (Fig. \ref{TestPhys}c).

We also compared the depletions with the molecular fraction
$f(\mathrm{H}_2) = 
2 N(\mathrm{H}_2) / [N(\ion{H}{i})+ 2 N(\mathrm{H}_2) ]$, which is
sensitive to physical conditions. 
Some evidence for a trend
 between differential depletions and $f(\mathrm{H}_2)$
has been reported in previous work (Levshakov et al. 2000; Ledoux et al. 2003).
The limited number
of measurements in common with the present sample are 
marginally consistent
with the existence of such a trend  (Fig. \ref{TestPhys}d). 
In any case, this trend would only
  confirm the  key role of dust
in the process of $\mathrm{H}_2$ formation, 
but would not provide evidence for a direct link
between the physical properties and the abundance of dust.  
In fact, the correlation is expected  because
 dust acts as a catalyst of $\mathrm{H}_2$ formation and provides a shield
against photo-dissociation of molecules by ultraviolet radiation.

 In summary,  there is no evidence that the evolution  of the
dust fractions/depletions is primarily governed  by variations of the physical
state of the gas.  In the following we consider   the
chemical evolution    as the main driver  of
the evolution of the dust abundance in DLA systems. 
However, we   expect that  variations of the physical state of the gas
will contribute to the   scatter
of dust abundance observed at a given metallicity.

\subsection{The dependence on chemical evolution}

The regular   trend shown by the data in  Fig. \ref{DepMet} 
is an argument in favour of a direct dependence of
depletions on the metallicity, i.e. on the chemical evolution. 

The  rise of the dust-to-metal ratio $f_\mathrm{Fe}$ in the course
of chemical evolution is unexpected,  given    
the approximately constant dust-to-metal ratio 
$\cal D/Z$ measured in local galaxies (Issa et al. 1990; Schmidt \& Boller 1993). 
However, the two results are not in contradiction since
the data in  Fig. \ref{DepMet} 
suggest the existence of   a flattening of    
$f_\mathrm{Fe} \approx {\cal D/Z}$     at [Fe/H] $\gsim -1$ dex, i.e.
when the metallicity approaches values typical
of the local sample of $\cal D/Z$ measurements.   

Also the   scatter of  $f_\mathrm{Fe}$ 
  in DLA systems  is consistent with
the  observed properties  of the $\cal D/Z$ ratio in present-day galaxies.
In fact, a spread of the $\cal D/Z$ ratio
is found when different types of   galaxies are compared,
such as the blue compact dwarfs and dwarf irregulars 
investigated by Lisenfeld \& Ferrara (1998). 
Therefore, the   scatter of    $f_\mathrm{Fe}$  
may simply reflect the inhomogeneity of 
    the population of DLA systems, which in fact is believed
to include galaxies of different types. 
The  $\cal D/Z$ ratio about 1/4 to 1/7 of the Galactic one recently found in  
a dwarf galaxy with an average metallicity $\simeq 1/4$ solar
(Galliano et al. 2003) is a further example indicating
that the properties of the dust fraction $f_\mathrm{Fe}$
in DLA systems are consistent
with the properties  of  the dust-to-metal ratios
$\cal D/Z$   in the local universe. 
Thefore, the trend seen in Fig. \ref{DepMet}  may represent
 a general characteristic of galactic evolution, rather than
 a specific property of DLA galaxies.
 
The   rise of  $\cal D/Z$ with metallicity  suggests that
{\em the efficiency of dust formation
is very low at the earliest stages of chemical evolution
but increases regularly as metals become
more abundant}. 
As a first step in 
interpreting this effect, we consider the possibility
that the most common processes of dust formation  may
show a dependence on the metallicity. 
Dust can be formed in the ejecta of
type II SNe (Moseley et al. 1989; Elmhamdi et al. 2003; Dunne et al. 2003)
and in the cool winds of late-type giant stars 
(see Sedlmayr 1989 and refs. therein).
It can probably also be formed in the ejecta of type Ia SNe, even though in this case the evidence is not direct  (Clayton et al. 1997). 
Other processes of dust formation are believed to give a minor
contribution on a galactic scale (see e.g. Dwek 1998). 
The composition of the dust is poorly known on observational grounds
and often inferred from chemical
equilibrium condensation calculations.
%
The relative abundance of iron in dust grains is not known, even though
iron could be present both in carbon-rich dust 
(e.g. Fe$_3$C; Lattimer et al. 1978), and in silicate
grains (e.g.  FeSiO$_3$, Fe$_2$SiO$_4$; Ossenkopf et al. 1992);
the dust formed by SNe Ia, if any, is expected to be iron-rich.  
In the following we consider the possible dependence on the metallicity 
of the main mechanisms of dust formation, assuming that
 a significant fraction of iron   can be incorporated in dust form
 in such processes.

Given the short life-times of their massive progenitors,
SNe II probably represent the only
source of dust at very high redshift. In fact, the time-scale for
dust production  by the winds of red giants or by SNe Ia is
of the order of $1$ Gyr, i.e. larger than the age
of the universe at $z \ga 4$. 
The existence of dust at $z > 4$ (e.g. Carilli et al. 2000) is a stringent
argument in favour of the production of dust by SNe II in the early universe. 
The systems with highest redshift in our sample are seen at about
$\ga 1.8$ Gyr after the start of the
reionization epoch (Fig. \ref{DepTime}a)
and a significant part of their dust
 must have been produced   by SNe II. 
 A  dust formation mechanism  for SNe II
dependent on metallicity has been 
found by Todini \& Ferrara (2001).  In this process
the density of heavy elements in the ejecta
becomes large enough at higher metallicities to allow the state of
supersaturation to be reached more easily, favouring the production of dust.
The metallicity-dependent dust formation in SNe II  
found by Todini \& Ferrara, together with the lack of
 other sources of dust, 
may explain the very low dust fraction  typical of
most of the highest-redshift systems  in our sample 
(shaded histogram in Fig. \ref{DepTime}b),
which have the lowest metallicities. 

After $\approx 1$ Gyr from the start of the chemical evolution
the dust starts to be injected also by the winds of red giants and
possibly by SNe Ia. We consider three possibilities:
(1)  these additional  contributions are negligible
compared  to the dust production of SNe II;
(2) the   production of dust in the wind
of late-type giants is dominant; and
(3) the production by SN Ia is dominant.

In the first case the metallicity-dependent dust production
of  SNe II may   explain the   rise of the dust fraction
with metallicity not only at the highest redshifts, but also at later stages
of evolution. 
 
 In the second case 
another mechanism may provide a metallicity-dependent
dust formation, namely  the 
{\em dependence on the metallicity of the wind efficiency}.
In fact, recent work 
on   dust production in giant stars indicates that  the 
efficiency of the stellar winds is lower at lower metallicity 
(Ferrarotti \& Gail 2003). Therefore,
the gradual rise of the efficiency of their winds 
with increasing metallicity may yield a gradual rise of the dust fraction.
However, this mechanism can explain the rise of $f_\mathrm{Fe}$ only
if iron is already present in the atmosphere of the   late-type giants,
since iron is not synthesized by these stars. 
This implies that this mechanism could only work at later stages
of evolution, $\approx 1$ Gyr {\em after} iron is already present in the gas out 
of which the future late-giants are born. 

In the third case,   the (hypothetical) 
iron-rich dust from SN Ia could provide an additional 
interpretation: the  {\em time delay} between the early production of  dust 
by SNe II and the subsequent production of iron-rich dust
by SNe Ia would work as a mechanism for increasing 
the iron dust fraction with time. 
A potential problem with this explanation is the possibility that
SNe Ia can only explode when [Fe/H] $\ga -1$ dex (Kobayashi et al. 1998), 
i.e. {\em after} the rapid increase of $f_\mathrm{Fe}$
seen in Fig \ref{DepMet}.  
 
In summary, all the important mechanisms of dust formation
might be able to provide a rise of the dust fraction in the
course of chemical evolution.  The most convincing mechanism, however,
is the metallicity-dependent production by SNe II proposed by Todini \& Ferrara (2001). 
Clearly, the evolution of the dust content will be determined not only
by the dust formation rate, but also by the destruction and accretion rates,
not discussed here  in detail. 
 By assuming that the ratio of  the destruction and accretion time-scales
 is proportional to the time-scale of star formation, Inoue (2003)
 predicts that the dust-to-metal ratio should increase
in the course of chemical evolution. 
The rise of the dust-to-metal ratio predicted by the "standard Galactic
model" of Inoue (solid curve in Fig. \ref{DepTime}a)
is less steep than the rise observed in DLA systems. 
The better agreement with the case of "no SNe destruction" 
(dotted line) may simply reflect the fact that iron is one of the most
refractory elements, not easily destroyed by the SNe shocks. 
The range of dust-to-metal ratios predicted
by  the three models shown in the Fig. 6 of Inoue (2003)
approximately bracket the observed data. 
The highest values of $f_\mathrm{Fe}$ observed
at  $\approx 2$ Gyr after the Big Bang suggest a very early epoch
of galaxy formation for these particular DLA systems, since  
even the model with the highest accretion rate (dashed line) has some
difficulty in reproducing these high values of $f_\mathrm{Fe}$,
even assuming a formation redshift  $z_f=20$. 

The above conclusions must be considered tentative given the
simplified nature of the analytical models and because
 the conversion between $f_\mathrm{Fe}$
and the dust-to-metal ratio  by mass is model-dependent
 (e.g. the conversion
requires some assumption on the dust composition). 
Nevertheless, this exercise is an example that shows how
the present data can be used to constrain   galactic
  evolution models incorporating the cycle of dust
formation and destruction. The implementation of the dust component
in realistic models calibrated on galaxies of the local universe can in turn be used to 
provide fresh clues on the redshift of formation and   the 
nature of DLA galaxies (e.g. Calura et al. 2003).

\section{Conclusions}

We have investigated
the evolutionary properties of the dust-to-metal ratio in DLA systems
with the aim of casting light on the early build-up of dust in galaxies. 
The results that we obtain are representative of 
high-redshift, metal-poor galaxies selected on the basis of their \ion{H}{i} absorption column density, i.e. $N(\ion{H}{i}) \geq 2 \times 10^{20}$ atoms cm$^{-2}$.

As an estimator of the dust-to-metal ratio in DLA systems
we have used the fraction of iron atoms in dust form,
$f_{\mathrm Fe}$. One advantage of this approach is that
iron is a refractory element which traces the presence of dust even in harsh
interstellar environments. 
The dust fraction  $f_{\mathrm Fe}$ has been derived  
by  comparing   the iron abundance  with the abundance
of the volatile element Zn.   
As in Paper I,   we take into account the fact
 that also Zn can be incorporated in dust form. 
However, the   methodology adopted here features  
important improvements over the one followed in Paper I: 
(1) the intrinsic Zn/Fe  ratio in DLA systems
is now a free input parameter;
(2) the dust chemical composition is allowed to vary   
in different galactic environments (see Paper II);
(3) we prefer  the dust-to-metal ratio, rather than the less
reliable dust-to-gas ratio, to study the evolution of the dust content. 
%
The main results  of the present   study 
 can be summarized as follows. 
 
 The frequency distribution of iron depletions, 
 $\delta({\mathrm Fe}) = \log (1-f_{\mathrm Fe})$, peaks at 
 $\delta({\mathrm Fe}) \sim 0$, and decreases regularly with 
 increasing $|\delta({\mathrm Fe})|$. 
 {\em A significant fraction of DLA systems has
 depletions lower than the lowest values measured in the Galactic ISM}, i.e. those
 of the Galactic halo gas. The frequently quoted claim 
 that depletions of DLA systems
 "are typical of Galactic warm gas" is correct only for the remaining systems.
 High values of depletion, typical of Galactic
 cold disk gas, are not found. 
 
 {\em The dust fraction $f_\mathrm{Fe}$  increases with
 metallicity}, albeit with a significant scatter and a
 {\em flattening
 of the trend when [Fe/H] $\ga -1$ dex}. 
 An empirical relation between dust fraction and metallicity  of the type
 $f_\mathrm{Fe} \approx 
f_{\mathrm{Fe},\circ} -10^{\alpha(\mathrm{[Fe/H]}-\mathrm{[Fe/H]}_{\circ})}$,
with $f_{\mathrm{Fe},\circ} \simeq 0.9$,
$\alpha \simeq -1.2$, and
$\mathrm{[Fe/H]}_{\circ} \simeq -1.7$,
provides a good fit to the data. 
 
%

The comparison between the dust fraction  $f_\mathrm{Fe}$
and the look-back time, $t_l$
(derived from the absorption redshift), reveals a weak evidence
for a rise of $f_\mathrm{Fe}$ with cosmic time. 
In spite of a large scatter,  
the data are consistent with a fast rise of $f_\mathrm{Fe}$ starting at
about 2 Gyr after the Big Bang and lasting a few billion years. 
Indeed, the comparison of the frequency distributions of depletions    
for the two sub-samples with  $t_l \geq 10.5$ Gyr and $t_l < 10.5$ Gyr
suggests that the dust fraction does increase with cosmic time.  

The above conclusions are rather stable for different choices
of the input parameters required for the determination of $f_\mathrm{Fe}$.
 In particular, they are not affected by changes of the adopted guess of the 
intrinsic Zn/Fe ratio in DLA systems,  [Zn/Fe]$_\mathrm{g}$, in line
with the measurements of the same ratio in metal-poor stars. 
The large amount of measurements recently
reported by three independent groups consistently
indicate that the   Zn/Fe ratio in metal-poor stars
is very close to solar and probably constant in the metallicity interval 
typical of DLA systems. 
 In any case,  {\em assuming that the intrinsic Zn/Fe ratio
decreases with chemical evolution}, as reported by some authors, 
{\em the existence of the trends with metallicity and cosmic time
would even be reinforced}.

 We have tested the robustness of the results in light of
the selection bias which are known to affect the
measurements of the 
 \ion{H}{i}, \ion{Fe}{ii}, and \ion{Zn}{ii} column densities. 
From the analysis of a mock sample of DLA systems
with assigned values of dust fraction and metallicity
we conclude that the bias effects are small and therefore 
that 
 {\em 
 the trend between dust fraction and metallicity represents
 a genuine property of DLA systems. This is also true 
 for the distribution of depletions, which is peaked at low values 
  and shows a  
 deficiency of systems with depletions typical of Galactic cold disk gas. }
 The dust obscuration bias may in part be responsible for such
deficiency, but probably only for the systems of high metallicity.  

The presence of different interstellar phases with a range of physical
parameters in DLA systems may contribute to the scatter
of the dust fractions measured at each metallicity. 
However, the depletions  do not show correlations
with physical parameters
measured in DLA systems, such as the spin temperature
or the  \ion{C}{ii}/\ion{C}{i} ionization ratio. 
We propose that {\em the evolutionary trends
of the dust fraction in DLA systems  are  mostly governed
by the chemical evolution} of the galaxies where the absorptions
originate, {\em rather than by the evolution of their  physical properties}. 

The extremely low dust fraction of iron  measured at 
[Fe/H]$ \lsim -2$ dex suggests that dust formation is very inefficient
during the earliest stages of galactic chemical evolution.
We considered the possibility that 
the main mechanisms of dust formation  in galaxies
may have a metallicity-dependent efficiency.     
Indeed, previous   work already indicates that
the dust formation efficiency in the ejecta of SNe II and in
the winds of late-type giants may increase with metallicity.
We argue that, in addition, the injection of iron-rich dust
by type SNe Ia may yield a rise of $f_\mathrm{Fe}$ in the course of evolution. 
The metallicity-dependent dust formation in SN II seems to be the
most promising candidate for explaining the rise
of $f_\mathrm{Fe}$ at [Fe/H] $\la -1$ dex. 
 
In order to reproduce the observed evolution of the dust content of DLA systems
a simultaneous treatment of the main processes of
dust formation, accretion and destruction is required.  
The observed increase of $f_\mathrm{Fe}$  with metallicity
is contrary to the expectations
of current chemical evolution models, which postulate or infer an approximately
constant dust-to-metal ratio, with the exception of the simple,
analytical  models by Inoue (2003), which predict an increase of 
the dust-to-metal ratio $\cal{D/Z}$, starting from very low values. 
The present results may represent a unique guideline for  
incorporating    the dust component 
 in well-tested models, 
 particularly for the early stages of evolution.  
The successful implementation of the observed
trends between $f_\mathrm{Fe}$, metallicity and cosmic time
in realistic models
will provide fresh clues  on the formation redshift and
nature of DLA galaxies. 
 
 \begin{acknowledgements}
I wish to thank Andrea Ferrara,  Akio Inoue 
and Patrick Petitjean for their useful comments
on the original version of the manuscript.   
\end{acknowledgements}

\end{document}